\documentclass[twocolumn,letterpaper,aps,prc,superscriptaddress,longbibliography,nofootinbib,floatfix]{revtex4-1}

\usepackage{graphicx}	% Include figure files
\usepackage{multirow}
\usepackage{xspace}	% Include xspace

\newcommand{\pt}{\mbox{$p_T$}\xspace}

\newcommand{\snn}{\mbox{$\sqrt{s_{_{NN}}}$}\xspace}

\newcommand{\pp}{\mbox{$p$$+$$p$}\xspace}
\newcommand{\pdA}{\mbox{$p/d$$+$$A$}\xspace}
\renewcommand{\AA}{\mbox{$A$$+$$A$}\xspace}

\newcommand{\dau}{\mbox{$d$$+$Au}\xspace}
\newcommand{\heau}{\mbox{$^3$He$+$Au}\xspace}

\newcommand{\ppb}{\mbox{$p$$+$Pb}\xspace}
\newcommand{\pau}{\mbox{$p$$+$Au}\xspace}
\newcommand{\auau}{\mbox{Au$+$Au}\xspace}
\newcommand{\cucu}{\mbox{Cu$+$Cu}\xspace}
\newcommand{\pbpb}{\mbox{Pb$+$Pb}\xspace}

\newcommand{\piz}{\mbox{$\pi^0$}\xspace}

\newcommand{\gevc}{\mbox{GeV/$c$}\xspace}
\newcommand{\gevcc}{\mbox{GeV/$c^2$}\xspace}

\begin{document}

\title{Measurement of emission-angle anisotropy via long-range angular 
correlations with high-$p_T$ hadrons in $d$$+$Au and $p$$+$$p$ collisions 
at $\sqrt{s_{_{NN}}}=200$ GeV}

\newcommand{\abilene}{Abilene Christian University, Abilene, Texas 79699, USA}
\newcommand{\augie}{Department of Physics, Augustana University, Sioux Falls, South Dakota 57197, USA}
\newcommand{\banaras}{Department of Physics, Banaras Hindu University, Varanasi 221005, India}
\newcommand{\barc}{Bhabha Atomic Research Centre, Bombay 400 085, India}
\newcommand{\baruch}{Baruch College, City University of New York, New York, New York, 10010 USA}
\newcommand{\bnlcoll}{Collider-Accelerator Department, Brookhaven National Laboratory, Upton, New York 11973-5000, USA}
\newcommand{\bnlphys}{Physics Department, Brookhaven National Laboratory, Upton, New York 11973-5000, USA}
\newcommand{\caucr}{University of California-Riverside, Riverside, California 92521, USA}
\newcommand{\charlesczech}{Charles University, Ovocn\'{y} trh 5, Praha 1, 116 36, Prague, Czech Republic}
\newcommand{\chonbuk}{Chonbuk National University, Jeonju, 561-756, Korea}
\newcommand{\ciae}{Science and Technology on Nuclear Data Laboratory, China Institute of Atomic Energy, Beijing 102413, People's Republic of China}
\newcommand{\cns}{Center for Nuclear Study, Graduate School of Science, University of Tokyo, 7-3-1 Hongo, Bunkyo, Tokyo 113-0033, Japan}
\newcommand{\colorado}{University of Colorado, Boulder, Colorado 80309, USA}
\newcommand{\columbia}{Columbia University, New York, New York 10027 and Nevis Laboratories, Irvington, New York 10533, USA}
\newcommand{\czechtech}{Czech Technical University, Zikova 4, 166 36 Prague 6, Czech Republic}
\newcommand{\dapnia}{Dapnia, CEA Saclay, F-91191, Gif-sur-Yvette, France}
\newcommand{\debrecen}{Debrecen University, H-4010 Debrecen, Egyetem t{\'e}r 1, Hungary}
\newcommand{\elte}{ELTE, E{\"o}tv{\"o}s Lor{\'a}nd University, H-1117 Budapest, P{\'a}zm{\'a}ny P.~s.~1/A, Hungary}
\newcommand{\eszterhazy}{Eszterh\'azy K\'aroly University, K\'aroly R\'obert Campus, H-3200 Gy\"ongy\"os, M\'atrai \'ut 36, Hungary}
\newcommand{\ewha}{Ewha Womans University, Seoul 120-750, Korea}
\newcommand{\fit}{Florida Institute of Technology, Melbourne, Florida 32901, USA}
\newcommand{\fsu}{Florida State University, Tallahassee, Florida 32306, USA}
\newcommand{\gsu}{Georgia State University, Atlanta, Georgia 30303, USA}
\newcommand{\hiroshima}{Hiroshima University, Kagamiyama, Higashi-Hiroshima 739-8526, Japan}
\newcommand{\howard}{Department of Physics and Astronomy, Howard University, Washington, DC 20059, USA}
\newcommand{\ihepprot}{IHEP Protvino, State Research Center of Russian Federation, Institute for High Energy Physics, Protvino, 142281, Russia}
\newcommand{\illuiuc}{University of Illinois at Urbana-Champaign, Urbana, Illinois 61801, USA}
\newcommand{\inrras}{Institute for Nuclear Research of the Russian Academy of Sciences, prospekt 60-letiya Oktyabrya 7a, Moscow 117312, Russia}
\newcommand{\instpasczech}{Institute of Physics, Academy of Sciences of the Czech Republic, Na Slovance 2, 182 21 Prague 8, Czech Republic}
\newcommand{\isu}{Iowa State University, Ames, Iowa 50011, USA}
\newcommand{\jaea}{Advanced Science Research Center, Japan Atomic Energy Agency, 2-4 Shirakata Shirane, Tokai-mura, Naka-gun, Ibaraki-ken 319-1195, Japan}
\newcommand{\jyvaskyla}{Helsinki Institute of Physics and University of Jyv{\"a}skyl{\"a}, P.O.Box 35, FI-40014 Jyv{\"a}skyl{\"a}, Finland}
\newcommand{\kek}{KEK, High Energy Accelerator Research Organization, Tsukuba, Ibaraki 305-0801, Japan}
\newcommand{\korea}{Korea University, Seoul, 136-701, Korea}
\newcommand{\kurchatov}{National Research Center ``Kurchatov Institute", Moscow, 123098 Russia}
\newcommand{\kyoto}{Kyoto University, Kyoto 606-8502, Japan}
\newcommand{\labllr}{Laboratoire Leprince-Ringuet, Ecole Polytechnique, CNRS-IN2P3, Route de Saclay, F-91128, Palaiseau, France}
\newcommand{\lahorelums}{Physics Department, Lahore University of Management Sciences, Lahore 54792, Pakistan}
\newcommand{\lawllnl}{Lawrence Livermore National Laboratory, Livermore, California 94550, USA}
\newcommand{\losalamos}{Los Alamos National Laboratory, Los Alamos, New Mexico 87545, USA}
\newcommand{\lpc}{LPC, Universit{\'e} Blaise Pascal, CNRS-IN2P3, Clermont-Fd, 63177 Aubiere Cedex, France}
\newcommand{\lund}{Department of Physics, Lund University, Box 118, SE-221 00 Lund, Sweden}
\newcommand{\lyon}{IPNL, CNRS/IN2P3, Univ Lyon, Université Lyon 1, F-69622, Villeurbanne, France}
\newcommand{\maryland}{University of Maryland, College Park, Maryland 20742, USA}
\newcommand{\mass}{Department of Physics, University of Massachusetts, Amherst, Massachusetts 01003-9337, USA}
\newcommand{\michigan}{Department of Physics, University of Michigan, Ann Arbor, Michigan 48109-1040, USA}
\newcommand{\muenster}{Institut f\"ur Kernphysik, University of Muenster, D-48149 Muenster, Germany}
\newcommand{\muhlenberg}{Muhlenberg College, Allentown, Pennsylvania 18104-5586, USA}
\newcommand{\myongji}{Myongji University, Yongin, Kyonggido 449-728, Korea}
\newcommand{\nagasaki}{Nagasaki Institute of Applied Science, Nagasaki-shi, Nagasaki 851-0193, Japan}
\newcommand{\nara}{Nara Women's University, Kita-uoya Nishi-machi Nara 630-8506, Japan}
\newcommand{\natmephi}{National Research Nuclear University, MEPhI, Moscow Engineering Physics Institute, Moscow, 115409, Russia}
\newcommand{\newmex}{University of New Mexico, Albuquerque, New Mexico 87131, USA}
\newcommand{\nmsu}{New Mexico State University, Las Cruces, New Mexico 88003, USA}
\newcommand{\ohio}{Department of Physics and Astronomy, Ohio University, Athens, Ohio 45701, USA}
\newcommand{\ornl}{Oak Ridge National Laboratory, Oak Ridge, Tennessee 37831, USA}
\newcommand{\orsay}{IPN-Orsay, Univ.~Paris-Sud, CNRS/IN2P3, Universit\'e Paris-Saclay, BP1, F-91406, Orsay, France}
\newcommand{\peking}{Peking University, Beijing 100871, People's Republic of China}
\newcommand{\pnpi}{PNPI, Petersburg Nuclear Physics Institute, Gatchina, Leningrad region, 188300, Russia}
\newcommand{\riken}{RIKEN Nishina Center for Accelerator-Based Science, Wako, Saitama 351-0198, Japan}
\newcommand{\rikjrbrc}{RIKEN BNL Research Center, Brookhaven National Laboratory, Upton, New York 11973-5000, USA}
\newcommand{\rikkyo}{Physics Department, Rikkyo University, 3-34-1 Nishi-Ikebukuro, Toshima, Tokyo 171-8501, Japan}
\newcommand{\saispbstu}{Saint Petersburg State Polytechnic University, St.~Petersburg, 195251 Russia}
\newcommand{\saopaulo}{Universidade de S{\~a}o Paulo, Instituto de F\'{\i}sica, Caixa Postal 66318, S{\~a}o Paulo CEP05315-970, Brazil}
\newcommand{\seoulnat}{Department of Physics and Astronomy, Seoul National University, Seoul 151-742, Korea}
\newcommand{\stonybrkc}{Chemistry Department, Stony Brook University, SUNY, Stony Brook, New York 11794-3400, USA}
\newcommand{\stonycrkp}{Department of Physics and Astronomy, Stony Brook University, SUNY, Stony Brook, New York 11794-3800, USA}
\newcommand{\sungskku}{Sungkyunkwan University, Suwon, 440-746, Korea}
\newcommand{\tenn}{University of Tennessee, Knoxville, Tennessee 37996, USA}
\newcommand{\titech}{Department of Physics, Tokyo Institute of Technology, Oh-okayama, Meguro, Tokyo 152-8551, Japan}
\newcommand{\tsukuba}{Tomonaga Center for the History of the Universe, University of Tsukuba, Tsukuba, Ibaraki 305, Japan}
\newcommand{\vandy}{Vanderbilt University, Nashville, Tennessee 37235, USA}
\newcommand{\waseda}{Waseda University, Advanced Research Institute for Science and Engineering, 17  Kikui-cho, Shinjuku-ku, Tokyo 162-0044, Japan}
\newcommand{\weizmann}{Weizmann Institute, Rehovot 76100, Israel}
\newcommand{\wigner}{Institute for Particle and Nuclear Physics, Wigner Research Centre for Physics, Hungarian Academy of Sciences (Wigner RCP, RMKI) H-1525 Budapest 114, POBox 49, Budapest, Hungary}
\newcommand{\yonsei}{Yonsei University, IPAP, Seoul 120-749, Korea}
\newcommand{\zagreb}{Department of Physics, Faculty of Science, University of Zagreb, Bijeni\v{c}ka c.~32 HR-10002 Zagreb, Croatia}
\affiliation{\abilene}
\affiliation{\augie}
\affiliation{\banaras}
\affiliation{\barc}
\affiliation{\baruch}
\affiliation{\bnlcoll}
\affiliation{\bnlphys}
\affiliation{\caucr}
\affiliation{\charlesczech}
\affiliation{\chonbuk}
\affiliation{\ciae}
\affiliation{\cns}
\affiliation{\colorado}
\affiliation{\columbia}
\affiliation{\czechtech}
\affiliation{\dapnia}
\affiliation{\debrecen}
\affiliation{\elte}
\affiliation{\eszterhazy}
\affiliation{\ewha}
\affiliation{\fit}
\affiliation{\fsu}
\affiliation{\gsu}
\affiliation{\hiroshima}
\affiliation{\howard}
\affiliation{\ihepprot}
\affiliation{\illuiuc}
\affiliation{\inrras}
\affiliation{\instpasczech}
\affiliation{\isu}
\affiliation{\jaea}
\affiliation{\jyvaskyla}
\affiliation{\kek}
\affiliation{\korea}
\affiliation{\kurchatov}
\affiliation{\kyoto}
\affiliation{\labllr}
\affiliation{\lahorelums}
\affiliation{\lawllnl}
\affiliation{\losalamos}
\affiliation{\lpc}
\affiliation{\lund}
\affiliation{\lyon}
\affiliation{\maryland}
\affiliation{\mass}
\affiliation{\michigan}
\affiliation{\muenster}
\affiliation{\muhlenberg}
\affiliation{\myongji}
\affiliation{\nagasaki}
\affiliation{\nara}
\affiliation{\natmephi}
\affiliation{\newmex}
\affiliation{\nmsu}
\affiliation{\ohio}
\affiliation{\ornl}
\affiliation{\orsay}
\affiliation{\peking}
\affiliation{\pnpi}
\affiliation{\riken}
\affiliation{\rikjrbrc}
\affiliation{\rikkyo}
\affiliation{\saispbstu}
\affiliation{\saopaulo}
\affiliation{\seoulnat}
\affiliation{\stonybrkc}
\affiliation{\stonycrkp}
\affiliation{\sungskku}
\affiliation{\tenn}
\affiliation{\titech}
\affiliation{\tsukuba}
\affiliation{\vandy}
\affiliation{\waseda}
\affiliation{\weizmann}
\affiliation{\wigner}
\affiliation{\yonsei}
\affiliation{\zagreb}
\author{A.~Adare} \affiliation{\colorado} 
\author{C.~Aidala} \affiliation{\mass} \affiliation{\michigan} 
\author{N.N.~Ajitanand} \altaffiliation{Deceased} \affiliation{\stonybrkc} 
\author{Y.~Akiba} \email[PHENIX Spokesperson: ]{akiba@rcf.rhic.bnl.gov} \affiliation{\riken} \affiliation{\rikjrbrc} 
\author{H.~Al-Bataineh} \affiliation{\nmsu} 
\author{J.~Alexander} \affiliation{\stonybrkc} 
\author{M.~Alfred} \affiliation{\howard} 
\author{A.~Angerami} \affiliation{\columbia} 
\author{K.~Aoki} \affiliation{\kek} \affiliation{\kyoto} \affiliation{\riken} 
\author{N.~Apadula} \affiliation{\isu} \affiliation{\stonycrkp} 
% ...
\collaboration{PHENIX Collaboration}  \noaffiliation

\date{\today}

%------------------------------------------------------------------------------|

\begin{abstract}

%\linenumbers

We present measurements of two-particle angular correlations between
high-transverse-momentum ($2<p_T<11$ GeV/$c$) $\pi^0$ observed at
midrapidity ($|\eta|<0.35$) and particles produced either at forward
($3.1<\eta<3.9$) or backward ($-3.7<\eta<-3.1$) rapidity in \dau and
\pp collisions at $\sqrt{s_{_{NN}}}=200$ GeV. The azimuthal angle
correlations for particle pairs with this large rapidity gap in the
Au-going direction exhibit a characteristic structure that persists up to
$p_T{\approx}6$ GeV/$c$ and which strongly depends on collision
centrality, which is a similar characteristic to the hydrodynamical
particle flow in A+A collisions.  The structure is absent in
the $d$-going direction as well as in \pp collisions, in the
transverse-momentum range studied. The results indicate that the
structure is shifted in the Au-going direction toward
more central collisions, similar to the charged-particle pseudorapidity
distributions.

\end{abstract}

\maketitle

\section{Introduction}

Azimuthal anisotropy in the multiparticle production from high-energy 
nucleus-nucleus collisions has been the subject of a great deal of 
study.  These final-state momentum anisotropies are believed to be the 
result of both spatial anisotropies in the initial geometry and 
hydrodynamic-like behavior in the subsequent evolution of the medium. 
The final-state patterns that can be modeled this way are thus often 
referred to as flow-like correlations, for which a central 
characteristic is that the majority of produced light-flavor hadrons 
will exhibit correlations with the initial collision geometry.  The 
measurement of azimuthal correlations of particles with a large rapidity 
gap (e.g. $|\Delta\eta|>3$) is particularly useful to extract the signal 
of the true flow contribution.  The near-side enhancement of the 
long-range correlation function is often called a ``ridge" structure, 
where the large relative pseudorapidity cut suppresses other sources of 
angular correlations, such as resonance decays or jet fragmentation, 
that are usually confined within $|\Delta\eta|\approx3$.

Analysis of flow-like correlations with hydrodynamical models has 
provided strong evidence for the creation of the quark-gluon plasma 
(QGP) state in the high-energy collisions of large nuclei, such as \auau 
and \cucu at the Relativistic Heavy Ion Collider (RHIC), and \pbpb at 
the Large Hadron Collider (LHC)~\cite{Heinz:2013th,Ollitrault:2008zz}. 
Great interest was sparked when flow-like behavior was first 
observed in small collision systems, including high-multiplicity \pp and 
\ppb at the 
LHC~\cite{Khachatryan:2010gv,CMS:2012qk,Abelev:2012ola,Aad:2012gla,Aad:2014lta,Aaboud:2016yar} 
and \dau at RHIC~\cite{Adare:2013piz,Adare:2014keg,Adamczyk:2015xjc}. 
Previously, these systems had been regarded as control systems where 
only nonQGP effects would be present.  Since then, similar flow-like 
observations have also been made in other small systems, including \pau 
and \heau.  The debate continues over whether the QGP is actually being 
created in this class of collisions~\cite{Adare:2015gla,Aidala:2016vgl}, 
and even at lower \snn~\cite{Aidala:2017ajz,Aidala:2017pup}. Possible 
explanations of these observations include 
hydrodynamics~\cite{Bozek:2013uha,Bzdak:2013zma,Romatschke:2015gxa,Weller:2017tsr} 
and Color-Glass-Condensate (CGC) models~\cite{Dusling:2013qoz}. The 
hydrodynamic models include both initial and final state effects, while 
the CGC-motivated models are based mainly on physics present in the 
initial state.  Interestingly, the kinetic transport model 
{\sc ampt}~\cite{Lin:2004en} also reproduces the observed flow structure 
fairly well~\cite{Aidala:2016vgl,Aidala:2017ajz,Aidala:2017pup}.
Similarly to hydrodynamics, \textsc{ampt} can translate the initial geometry into 
final-state momentum anisotropy, but via a very different mechanism, namely 
the anisotropic probability of partons to escape the partonic scattering stage~\cite{He:2015hfa}.

The PHENIX experiment has previously measured azimuthal correlations in 
\dau and \pp between charged particles produced at midrapidity 
(pseudorapidity $|\eta|\approx0$) and energy deposits in a forward 
calorimeter ($|\eta|\approx3.5$)~\cite{Adare:2014keg}. In those 
analyses, the reach in charged particle \pt was statistically limited to 
$p_T<3.5$\,\gevc. Measurements of azimuthal anisotropy at low \pt are
useful to study the collective behavior of the QGP medium. However,
at high \pt, azimuthal anisotropy signals can no longer be attributed
to the collective expansion of the bulk. Measurements in $p$$+$Pb at the 
LHC~\cite{Aad:2014lta,Chatrchyan:2012xq} have shown that $v_2$ decreases 
sharply between $4 \lesssim$ \pt $\lesssim 8$ GeV/$c$, reaching a small 
near-constant value above that point. It has been suggested that this 
high-\pt behavior might originate from jet quenching. Therefore, the 
present paper extends the measurements of two-particle correlations at 
RHIC to this kinematic region where nonhydrodynamic effects dominate.
We use the PHENIX high-energy photon trigger in the midrapidity region,
and explore mid-forward(backward) correlations in \dau and \pp up to
\pt~=~11\,\gevc with identified \piz at midrapidity.  

In large collision systems, the appearance of a near-side enhancement in 
azimuthal two-particle correlations is considered a hallmark signature 
of QGP collectivity.  Thus, early searches for collectivity in small 
collision systems focused on observing near-side enhancement.  However, 
unlike in $A$$+$$A$ collisions, elementary processes cannot be neglected 
when analyzing small systems.  Thus, even if collectivity exists, it may 
not be necessarily observed as a near-side enhancement because the ratio 
of quadrupole to dipole contributions is decreasing with multiplicity. 
This is particularly true for \pp and peripheral \dau collisions, as the 
``smallest" of the small systems considered in the present analysis. In 
light of this, the paper presents a wealth of data and attempts to 
characterize the shape of the two-particle correlation functions by 
investigating the behavior of the coefficients of the Fourier series 
fit, in relation to the appearance of a near-side enhancement.

In addition to measuring flow by the correlation of individual particles 
to the reaction plane, it is also possible to measure flow by the 
correlation of two-particles to each other. The advantage of this method 
is that one does not have to determine the reaction plane. If we write 
the azimuthal angle distribution of two particles $A$ and $B$, which are 
correlated to a reaction plane as:
\begin{eqnarray}
\frac{dN^A}{d\phi^A} &\propto& 1+\sum_{n} 2v_{n}^{A} \cos (n(\phi^A-\Psi_n)) \\
\frac{dN^B}{d\phi^B} &\propto& 1+\sum_{n} 2v_{n}^{B} \cos (n(\phi^B-\Psi_n)) 
\end{eqnarray}
then the azimuthal angle distributions for the two particle correlations
can be written as:
\begin{eqnarray}
\frac{dN^{AB}}{d\phi^{AB}} &\propto& 1+\sum_{n} 2v_{n}^{A}v_{n}^{B} \cos (n(\phi^A-\phi^B)) \\
&\equiv& 1+\sum_{n} 2c_{n} \cos (n(\phi^A-\phi^B)) 
\end{eqnarray}
Instead of measuring $v_n$, this paper presents measurements of 
$c_n$, the coefficient of the Fourier fit to the correlation functions,
because the factorization $c_n=v_{n}^{A}v_{n}^{B}$
holds only at low \pt, where the two particles are correlated with the
same event plane~\cite{Tannenbaum:2012ma}. This relation breaks down
when considering high \pt particles that are coming from the nonflow
contributions such as jet fragmentation.

%==================================
%phenix 28 phenix 28 phenix 28
%
% Start of section "Experiment and Dataset"
%
% "Analysis" made new section below, PWS 4/1
%

%---------------------------------------------------  Fig_1
\begin{figure*}[tbp]
\centering
\includegraphics[width=0.99\linewidth]{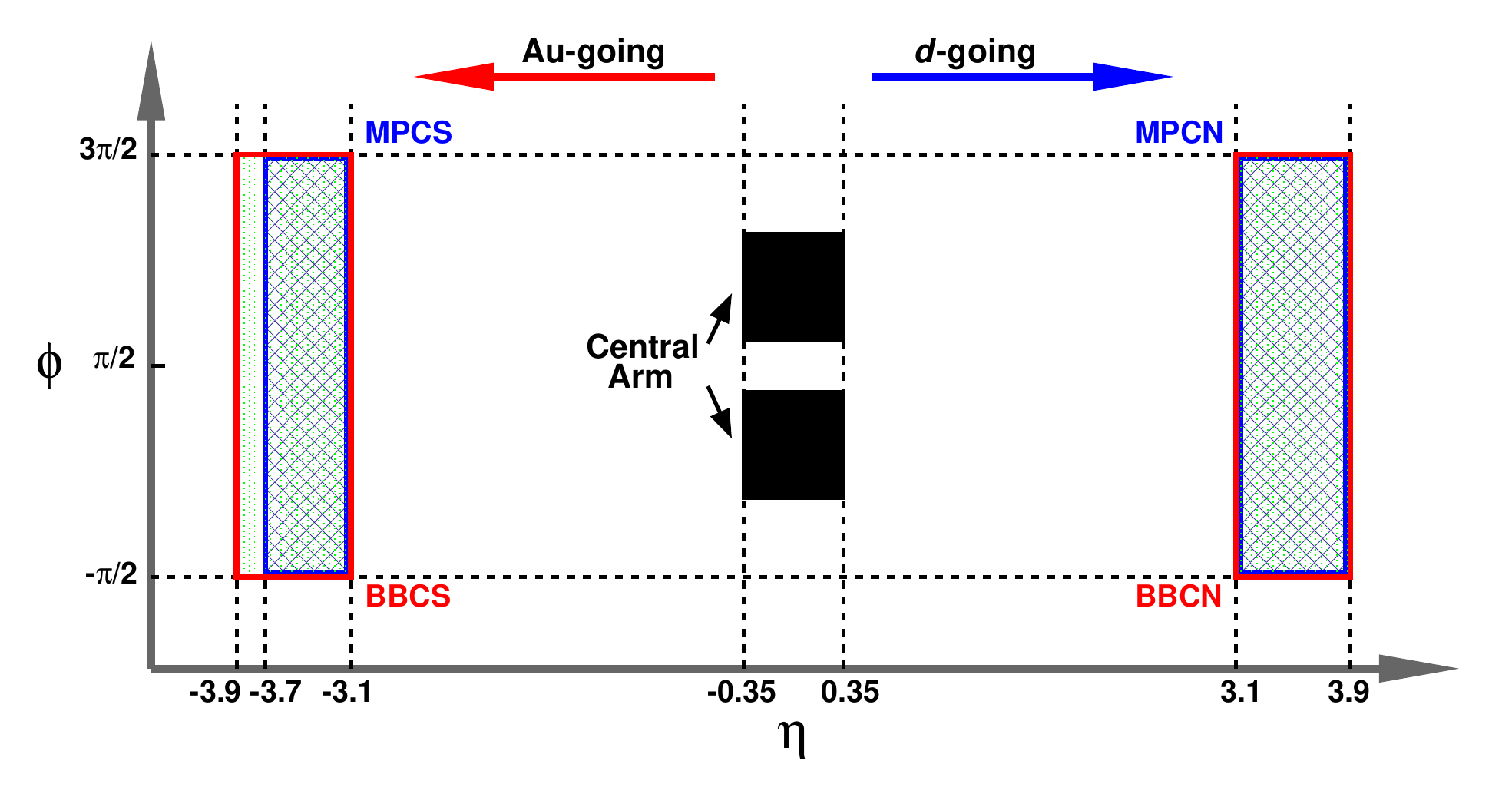}
\caption{Configuration in azimuth and pseudorapidity ($\phi$-$\eta$) 
coordinates of the PHENIX detector subsystems used in this analysis.  
The BBC and MPC detectors each cover $2\pi$ in azimuth in the forward 
and backward directions, while the two PHENIX Central Arms each subtend 
$\pi/2$ in azimuth.
}
\label{fig1}       
\end{figure*}

\section{Experiment and Dataset 
\label{ExpSection}}

A detailed description of the PHENIX detector system can be found 
elsewhere~\cite{Adcox:2003zm}. The principal detectors used in this 
analysis are the beam-beam counters (BBC), the muon-piston calorimeter 
(MPC) and the electromagnetic calorimeter (EMCal).  The BBCs are located 
north (BBCN, 3.1$<\eta<$3.9, $d$-going) and south (BBCS, 
-3.9$<\eta<$-3.1, Au-going) of the interaction point, covering the full 
azimuth and are sensitive to charged particles. In \dau collisions, 
the Au ions are accelerated in the Au-going direction.  The MPCs, 
which are high resolution electromagnetic calorimeters, are also located 
north (MPCN, 3.1$<\eta<$3.9) and south (MPCS, -3.7$<\eta<$-3.1) of the 
interaction point, in front of the BBCs, and cover the full azimuth. The 
south (north) MPC comprise 192 (220) PbWO$_4$ crystal towers with 
20.2 $X_0$ or 0.89 $\lambda_{\rm I}$~\cite{Adare:2013ekj}. The EMCal is 
located in the central (CNT) arms with pseudorapidity range $|\eta|<$0.35 
and covering two $\pi/2$ segments of the full azimuth.  
Figure~\ref{fig1} shows the acceptance of each relevant PHENIX detector 
subsystem in $\phi$-$\eta$ coordinates.

The \dau and \pp collision data used in this analysis were recorded 
in 2008 at RHIC.  The events triggered by a high energy deposit in a 4x4 
tower region of the EMCal in coincidence with the minimum bias (MB) 
requirement were selected in both the \pp and \dau data sets. The MB 
trigger was defined as the coincidence of at least one hit in the BBCS 
and BBCN.  A z-vertex cut of $|z|<~$30\,cm is applied, using the vertex 
calculated from the BBC timing information. The energy threshold of the 
4x4 towers is set to be 2.8\,GeV, however, due to the energy smearing 
effect, the towers also sample hits with lower energies but with lower 
efficiency.  The number of recorded events was $2.85{\times}10^8$ 
($9.64{\times}10^{10}$ MB equivalent) for the \pp and 
$6.51{\times}10^8$ ($1.40{\times}10^{11}$ MB equivalent) for the \dau 
collisions, which made it possible to measure the \piz-triggered 
long-range correlations up to \pt=11\,\gevc. In the case of \dau 
collisions, centrality was defined by the total charge deposited in BBCS 
(Au-going direction). Seven partially overlapping centrality bins have 
been considered, from the most central (0\%--5\%) to the most peripheral 
(60--88\,\%) collisions~\cite{Adare:2013nff}.

%===========================================================
%===========================================================
%
% Start of section "Analysis"
%
% New section divider added, PWS 4/17
%
\section{Analysis \label{AnaSection}}

The long-range two-particle correlation functions are constructed 
by pairing a high \pt  \piz   (``trigger" particle) found in the PHENIX 
EMCal with the energy deposit $E_{{\rm dep}}$ in each tower 
of one of the MPCs (``associated" hit).  In the following sections we
describe (i)~the \piz identification, (ii)~construction of the initial
azimuthal correlation functions, (iii)~correction for combinatoric
background in the \piz sample, and (iv)~fitting the corrected
correlation functions with a harmonic expansion.
Throughout this paper the results for central-MPC south (CNT-MPCS) and 
central-MPC north (CNT-MPCN) correlations are shown separately.

%========================
% 
%  Start of sub-section on pi-0 selection  added PWS 7/17
%

\subsection{\piz selection}

%%====================
%
% Figure with example photon pair mass spectrum
%

%---------------------------------------------------  Fig_2
\begin{figure}[htbp]
\includegraphics[width=1.0\linewidth]{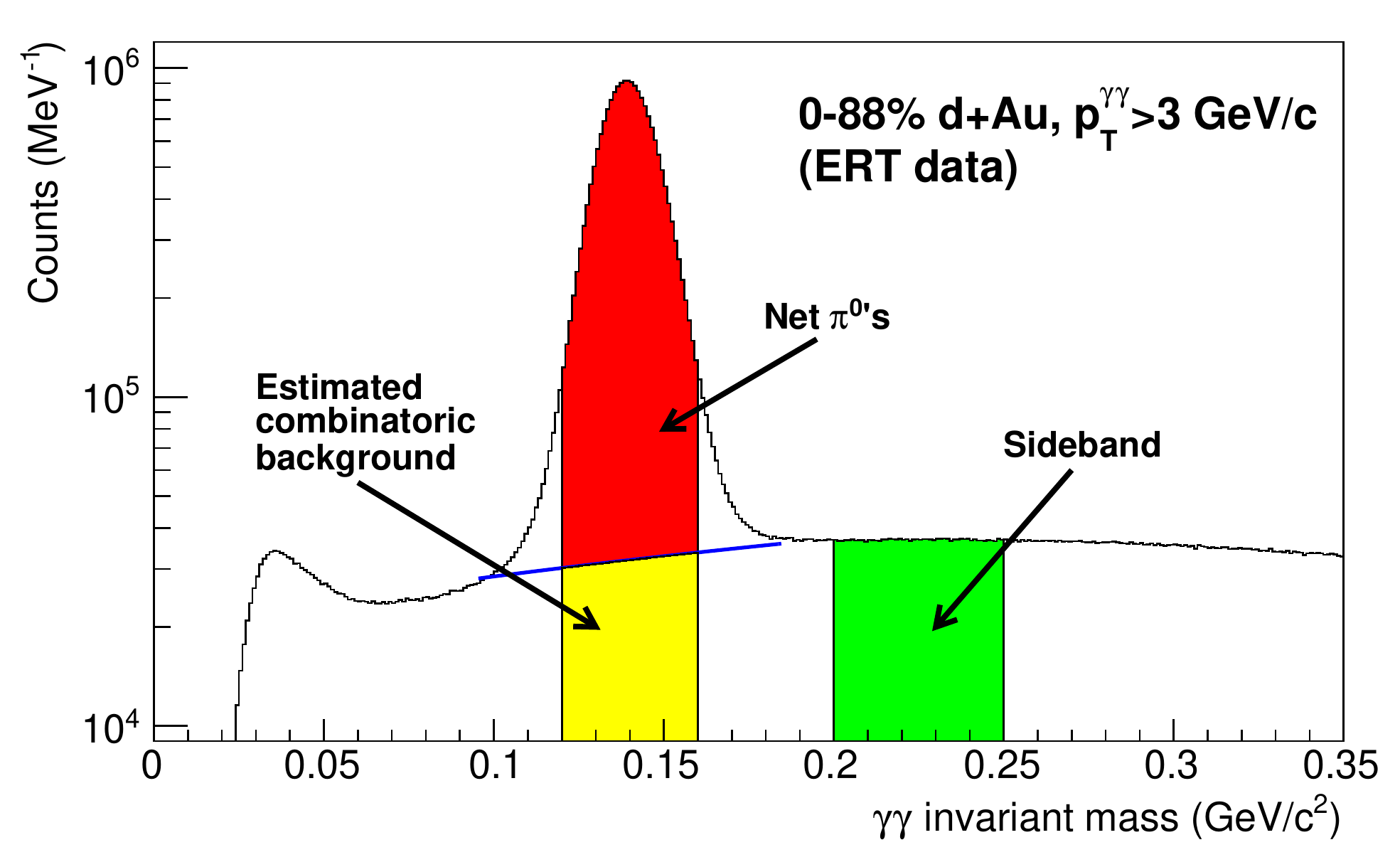}
\caption{
Invariant mass distribution for $\gamma\gamma$ pairs from \dau collisions 
as measured in the PHENIX central arm EMCal.  The [red] shaded ``Net \piz" 
peak is clearly visible above a small [yellow] shaded ``Estimated 
combinatoric background" in the same mass window 
$0.12<m_{\gamma\gamma}<0.16$\,\gevcc (note the semilog scale).  We 
estimate the combinatoric background by interpolating linearly between 
two points outside the peak, as shown by the [blue] line, which is 
obtained by fitting around the peak with a combined Gaussian and linear 
function. The purely combinatoric pairs in the shaded [green] ``Sideband" 
region are used to correct the correlation functions for the effects of 
background pairs in the peak region (see 
Section~\ref{subsec:sideband_correction}).
}
\label{Fig:InvMassDemo}
\end{figure}

Each trigger \piz was measured in the EMCal via the $\pi^0 \rightarrow 
\gamma\gamma$ decay channel using photon showers reconstructed using the 
standard PHENIX 
method~\cite{Adler:2006bw,Afanasiev:2009aa,Adare:2012wg}. The photon 
showers were identified using a shower-shape 
cut~\cite{Aphecetche:2003zr}.  A cut on the energy asymmetry of the 
photon pair $\alpha=|E_1-E_2|/(E_1+E_2)<0.7$ has been applied to 
reduce the combinatoric background.  A sample $\gamma\gamma$ invariant 
mass plot is shown in Fig.~\ref{Fig:InvMassDemo} for pairs with pair 
$\pt > 3$\,\gevc. The \piz mass region was defined as 
$0.12<m_{\gamma\gamma}<0.16$\,\gevcc, and every measured pair in this 
range was used in compiling the initial correlation functions, binned 
according to pair \pt.

%%===================
%
% Figure with photon pair background/signal ratio
%

%---------------------------------------------------  Fig_3
\begin{figure}[htbp]
\includegraphics[width=1.0\linewidth]{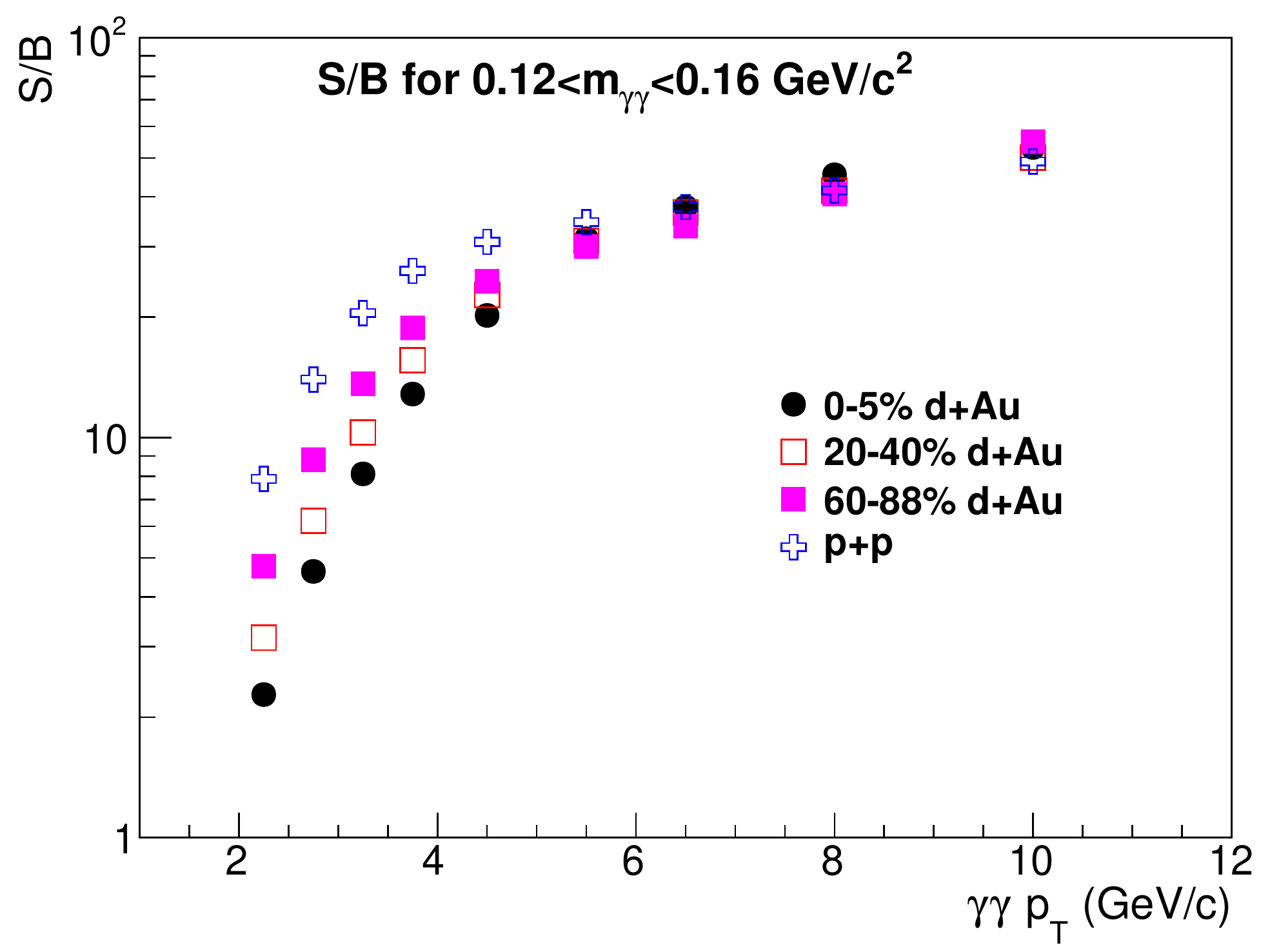}
\caption{Centrality and $\gamma\gamma$ \pt dependence
of the signal to combinatoric background ratio ($S/B$)
for pairs in the 
0.12$<m_{\gamma\gamma}<$0.16\,\gevcc mass window.}
\label{Fig:BBSRatio}
\end{figure}

As shown in Fig.~\ref{Fig:InvMassDemo} the \piz peak is quite prominent 
in the pair mass spectrum, on top of a small background continuum due 
primarily to combinatoric pairs. We estimated the level of this 
background in terms of the signal/background ratio $S/B$ within the 
chosen \piz mass window as shown in Fig.~\ref{Fig:BBSRatio}. The ratio 
was used for subtracting the combinatoric background contribution in the 
correlation functions as explained in the 
section~\ref{subsec:sideband_correction}.

%======================================
% 
% sub-section on constructing initial correlation function added PWS 7/17 
%

\subsection{Initial correlation functions}
\label{subsec:initial_corr_functions}

The procedure used to construct the initial \piz-MPC correlation 
functions is essentially the same as was used in our earlier analysis of 
central-arm charged track -- MPC correlations in \dau and \pp 
collisions~\cite{Adare:2014keg}. Over a selected event sample and \piz 
\pt bin, we compile the relative azimuthal angle distribution, 
$S(\Delta\phi,p_T)$, between $\gamma \gamma$ pairs in a given mass 
window and MPC towers in the same event

\begin{equation}
  S(\Delta\phi,p_T)= 
   \frac{ d(w_{{\rm tower}} N^{{\rm \gamma \gamma}(p_T){\rm -tower}}_{{\rm Same \; event}}) }{ d\Delta\phi},
\label{eq:CF_numerator}
\end{equation}
 
\noindent where $\Delta\phi$ = $\phi_{\gamma \gamma}-\phi_{{\rm tower}}$ 
is the azimuthal opening angle between the $\gamma \gamma$ pair-sum 
momentum direction and a line to the center of the MPC tower.  We choose 
the weighting for each tower to be the transverse energy 
$w_{{\rm tower}} = E_{{\rm dep}} \sin{(\theta_{{\rm tower}})}$, where 
$E_{{\rm dep}}$ is the energy deposit in that tower, and 
$\theta_{{\rm tower}}$ is the angular position of the tower with 
respect to the beam line. The $w_{{\rm tower}}$ introduces a \pt 
spectrum weight on the hit frequency in the MPC. The MPC towers with 
deposited energy $E_{{\rm dep}}>$ 0.3\,GeV were selected to avoid the 
background from noncollision noise sources ($\approx$75\,MeV) and to cut 
out the deposits by minimum ionizing particles ($\approx$245\,MeV).  To 
maximize statistics the energy is lowered compared to the one used in a 
previous publication~\cite{Adare:2014keg}.

In addition to physical pair correlations from the collisions, the shape 
of the same-event distribution $S(\Delta\phi,\pt)$ will reflect the 
effects of detector acceptance, detector inefficiencies, and kinematic 
cuts. We estimated these instrumental effects by constructing a 
mixed-event distribution $M(\Delta\phi,\pt)$ 
[Eq.~\ref{eq:CF_numerator}], but using $\gamma \gamma$ pairs from one event 
and MPC towers from a different event in the same event class 
(centrality and \piz \pt).  We then correct for instrumental effects by 
constructing the correlation function $C^{X}(\Delta\phi,\pt)$, for any 
particular choice $X$ of $\gamma\gamma$ pair selection criterion

\begin{equation}
C^{X}(\Delta\phi,p_T) = \frac{S^{X}(\Delta\phi,p_T)}{M^{X}(\Delta\phi,p_T)} 
\frac{\int M^{X}(\Delta\phi,p_T) \; d\Delta\phi}{\int S^{X}(\Delta\phi,p_T) 
d\Delta\phi}
\label{eq:CFX_definition}
\end{equation}

Both the same-event numerator and the mixed-event denominator have been 
normalized by their respective integrals.

%========================
% 
%  Start of sub-section on sideband correction  added PWS 7/17
%

\subsection{Combinatoric sideband correction}
\label{subsec:sideband_correction}

The initial correlation function is constructed using all pairs in the 
\piz mass window, which necessarily includes an admixture of both true 
\piz pairs and background pairs. Therefore, it will not reflect simply 
the true \piz-MPC correlation but rather a weighted average of the 
correlations of true \piz pairs and those of background pairs. Though 
the background is typically a small fraction of the signal, as shown in 
Fig.~\ref{Fig:BBSRatio}, we carried out the following correction to 
remove any influence from the background pairs.

We denote the initial correlation function constructed using all photon 
pairs in the \piz mass peak region as $C^{S+B}(\Delta\phi,\pt)$, because 
it contains correlations from both signal and background pairs. We then 
approximate the correlation function $C^{B}(\Delta\phi,\pt)$ that would 
result from using the background pairs only, by constructing a 
correlation function according to Eq.~\ref{eq:CFX_definition}, but with 
pairs chosen from the ``sideband'' mass region 
$0.20<m_{\gamma\gamma}<0.25$\,\gevcc [see Fig.~\ref{Fig:InvMassDemo}]. 
We then derive the true \piz-MPC correlation 
function $C(\Delta\phi,\pt)$, which would result from including only the 
true \piz decay pairs, by inverting the weighted average via

\begin{equation}
C^{}(\Delta\phi,p_T) =
\left(1+\frac{B}{S}\right) C^{S+B}(\Delta\phi,p_T)
-\frac{B}{S}C^{B}(\Delta\phi,p_T)
\label{eq:side_band_formula}
\end{equation}

\noindent where $B/S$ is the background-to-signal ratio in the peak 
region, which is the reciprocal of the number shown in 
Fig.~\ref{Fig:BBSRatio}. In practice, this correction for background 
pairs is very small; it does not change the harmonic amplitudes of the 
correlation function (see Section~\ref{subset:harmonic_expansion}) by 
more than a few percent of their value in the lowest $S/B$ cases and 
becomes negligible as $S/B$ increases toward higher \pt.

%---------------------------------------------------  Fig_4
\begin{figure*}[tbp]
\includegraphics[width=0.8\linewidth]{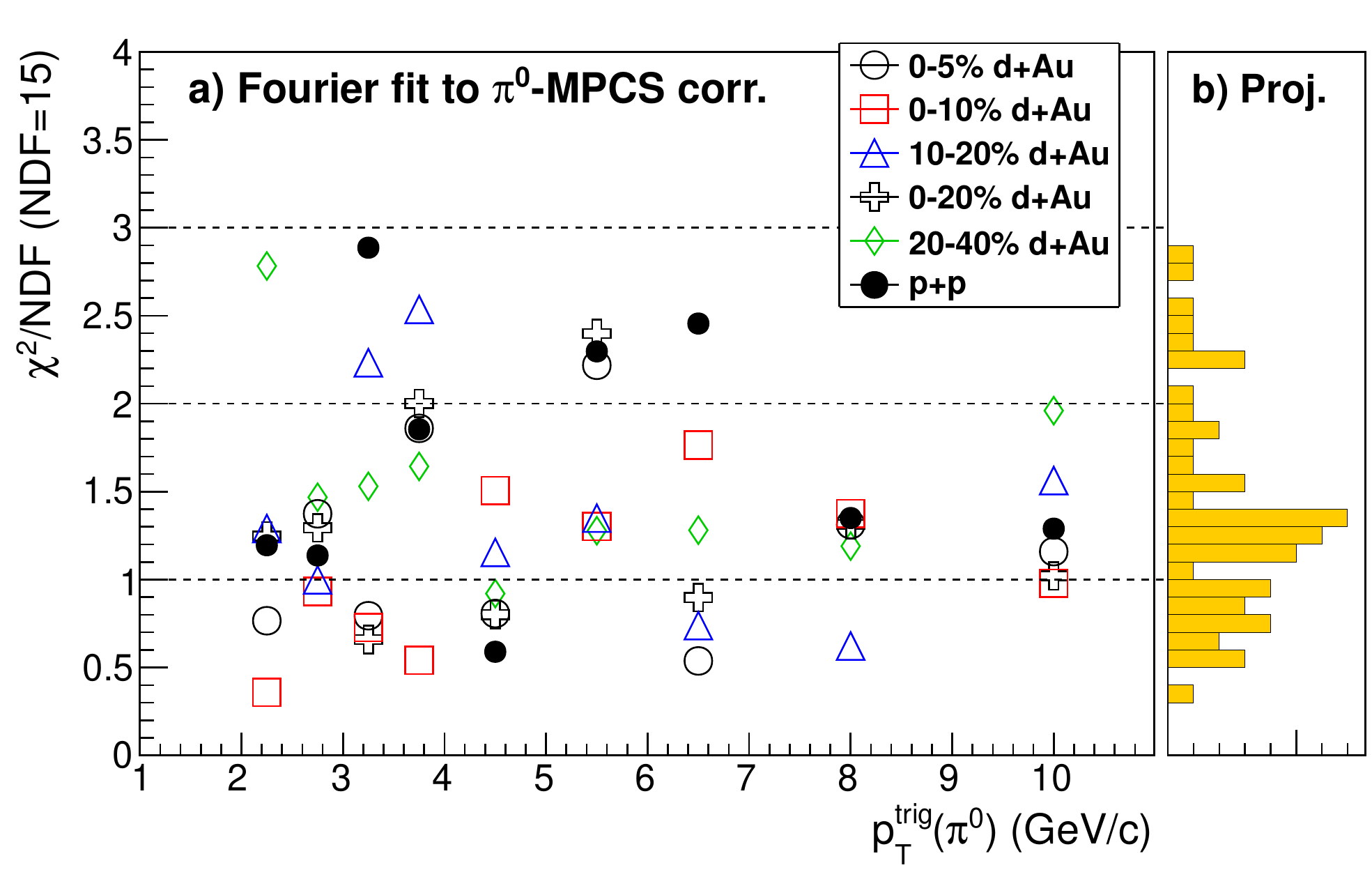}
\caption{(a) Goodness-of-fit parameter $\chi^2/$NDF for the harmonic 
fits in Eq.~\ref{eq:four_fit} to the corrected \piz-MPCS correlation 
functions, for different centrality and \piz \pt selections, and (b) 
their projection to the $y$-axis.
}
\label{Fig:chi2ndf}
\end{figure*}

%---------------------------------------------------  Fig_5
\begin{figure*}[tbp]
\begin{minipage}[t]{0.49\linewidth}
\includegraphics[width=0.99\linewidth]{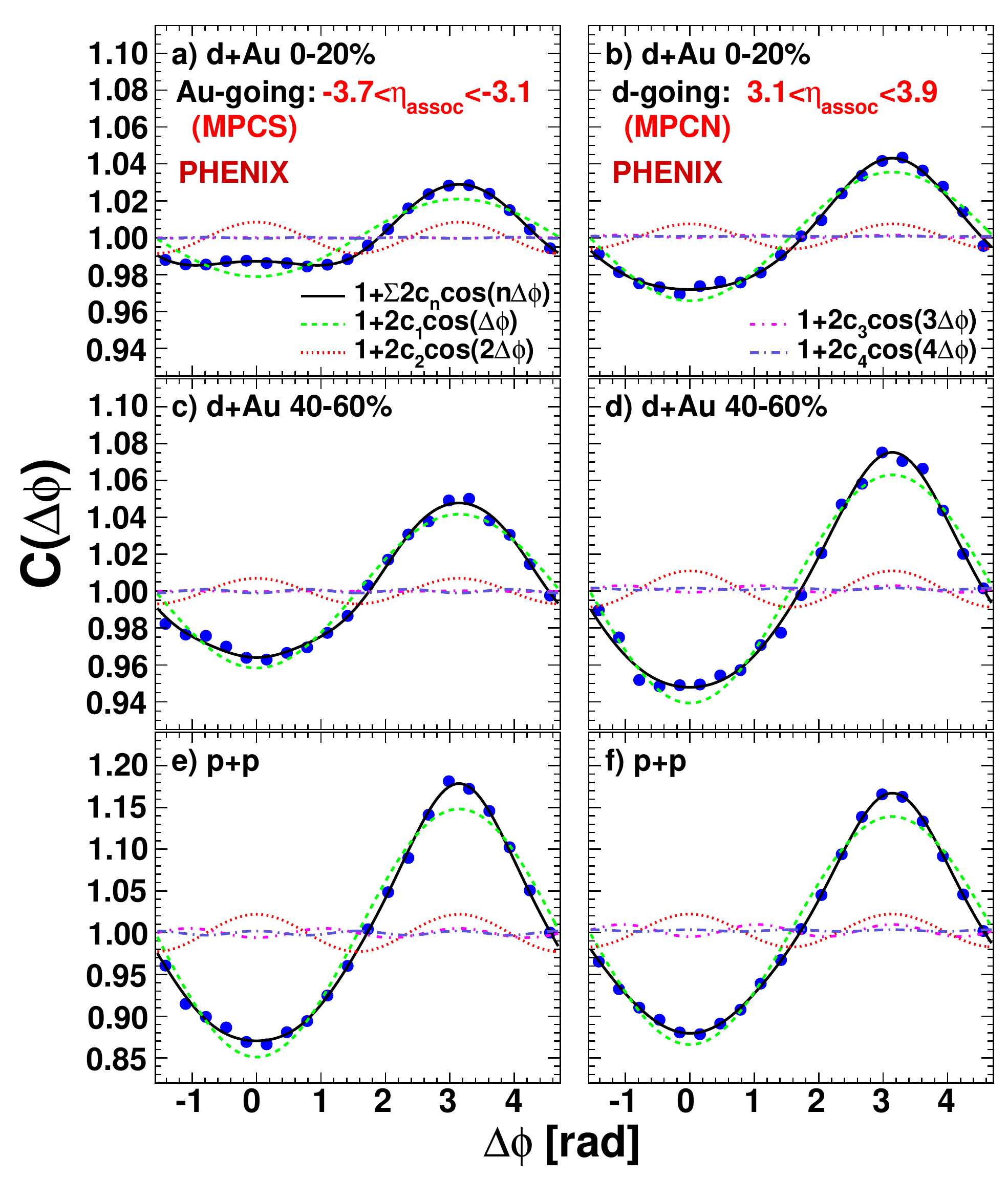}
\caption{Centrality dependence of correlation functions for \dau and 
\pp collisions at \snn=200\,GeV for $\pi^0$ in 
$|\eta_{\rm trig}|<0.35$ (CNT). 
(a, c, e) $\pi^0$ are associated with the towers in MPCS (Au-going 
direction) and (b, d, f) MPCN ($d$-going direction), for 
3$<p_T<$3.5\,\gevc.
}
\label{Fig:Pi0MPCSummary_1}
\end{minipage}
%---------------------------------------------------  Fig_6
\begin{minipage}[t]{0.49\linewidth}
\includegraphics[width=0.99\linewidth]{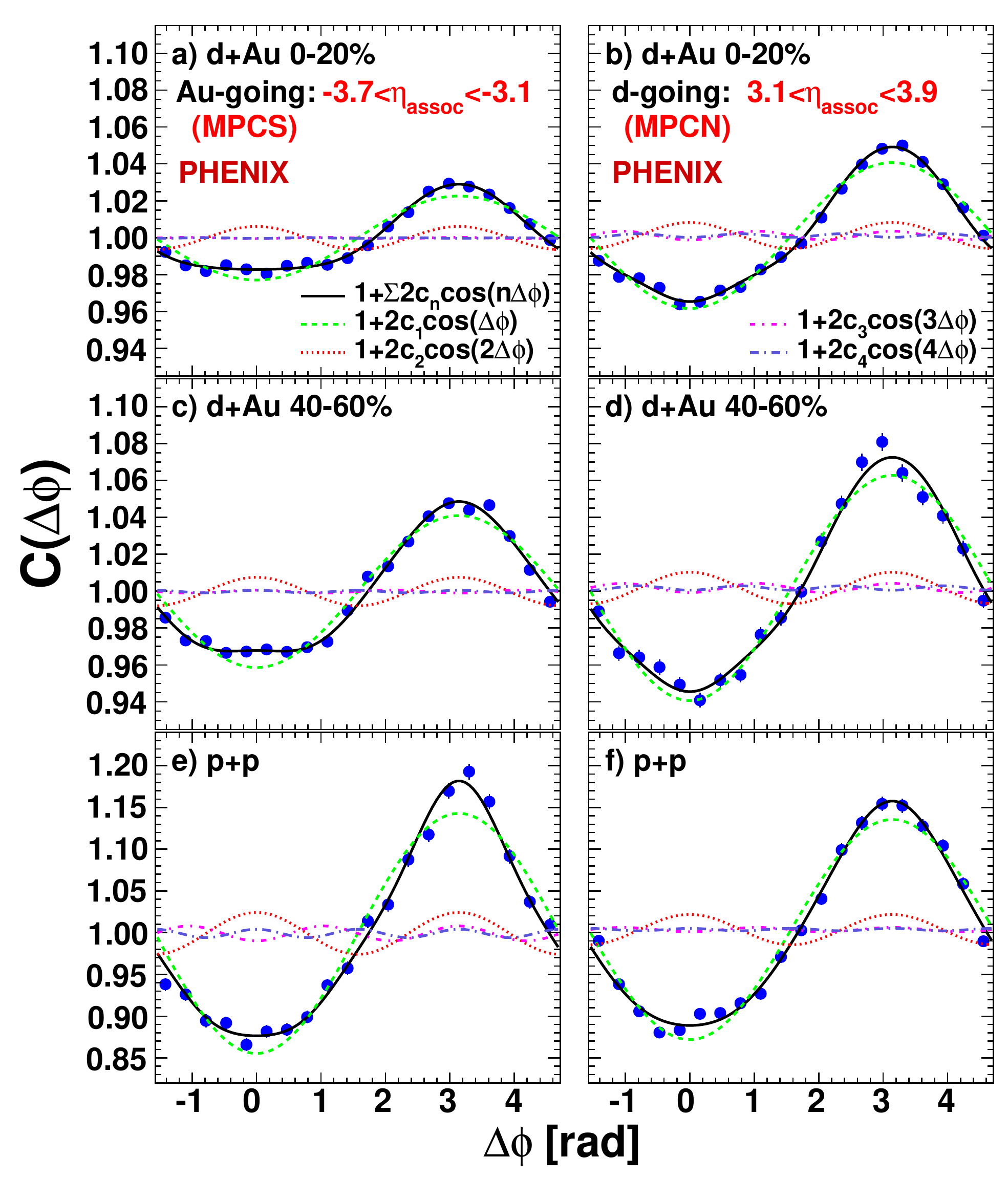}
\caption{The same as Fig.~\ref{Fig:Pi0MPCSummary_1}, except for 
5$<p_T<$6\,\gevc.
}
\label{Fig:Pi0MPCSummary_2}
\end{minipage}
\end{figure*}

%========================
% 
%  Start of sub-section on harmonic fit  added PWS 7/17
%

\subsection{Harmonic expansion fitting}
\label{subset:harmonic_expansion}

Our objective in this analysis is to examine the shapes of the \piz-MPC 
correlation functions across \piz \pt and collision system centrality 
classes. We quantify each correlation function by fitting them to an 
expansion in Fourier terms over $\Delta\phi$ up to fourth order via

\begin{equation}
C(\Delta\phi,p_T)=B_{0} \left( 1 + \sum_{n=1}^{4} 2 \, c_{n}(p_T) \cos(n \Delta\phi) \right)
\label{eq:four_fit}
\end{equation}

The fits were optimized using only the statistical errors
in the final correlation functions. The fit for each \pt and event class 
combination has five parameters: the four $c_{n}$ and an overall 
normalization.  Each correlation function was compiled in 20 bins of 
$\Delta\phi$, leaving 15 degrees of freedom (NDF) for each fit. 
The $C(\Delta\phi,\pt)$ with fit functions are shown in 
section~\ref{ResultAndDiscuss} and in the Appendix. The $\chi^2$/NDF 
goodness-of-fit values are compiled and shown in Fig.~\ref{Fig:chi2ndf}. 
There is no particular structure seen with \piz \pt or event class, and 
the distribution agrees with what would be expected for a $\chi^2$ 
estimator.

%%====================================
%
%  Figure with chi-square/NDF for harmonic fits
%

When we fit the correlation functions with $c_2$ fixed to zero, the 
$\chi^2$/NDF's are found to be as high as $\approx$40 around \pt=3\,\gevc, 
and don't reach $\chi^2$/NDF$\approx$4 before \pt$\approx$6\,\gevc, for both 
0\%--5\% central \dau and \pp collisions. This shows that the 
correlation functions have a significant second-order component.

%========================
% 
%  Start of sub-section on harmonic fit  added PWS 7/17
%

\subsection{Estimation of systematic errors}
\label{subsec:systematic_errors}

The systematic uncertainties of the measurement have been estimated as 
follows. The width of the \piz extraction window as well as the location 
and width of the sideband have been varied in five different 
combinations as listed in Table~\ref{tab1}. Note that the case-0 
corresponds to the standard windows in this analysis.

%%%%%%%%%%%%%%%%%%%%%%%%%%%%%%%%%%%%%%%%%%%%%  Table_I
\begin{table}[htbp]
\caption{Combination of \piz extraction and sideband windows for 
estimating systematic errors. Note that the case-0 corresponds to 
the standard windows in this analysis.
}
\begin{ruledtabular} \begin{tabular}{ccc}
Case & \piz window (\gevcc) & sideband window (\gevcc) \\
\hline
0 & 0.12--0.16 & 0.20--0.25\\
1 & 0.12--0.16 & 0.25--0.30\\
2 & 0.12--0.16 & 0.06--0.09\\
3 & 0.12--0.16 & 0.06--0.09 + 0.20--0.30\\
4 & 0.10--0.18 & 0.20--0.25\\
5 & 0.13--0.15 & 0.20--0.25\\
\end{tabular} \end{ruledtabular}
\label{tab1}
\end{table}

In the sixth case the original windows were kept as case-0 but the 
asymmetry cut was changed to $\alpha<0.5$. Following the exact same 
procedure for obtaining the true \piz-MPC correlation functions as 
described in the previous sections, the correlation functions for the 
six cases were obtained and the values of $c_2$ and $-c_2/c_1$ were 
re-calculated. The deviations for the case-0 values, with respect to the 
standard result, were calculated and averaged over the six cases. 
The averaged deviations are the systematic uncertainties. The resulting 
uncertainties on $c_2$ are 2\% for \pp (all \pt), and for the 0\%--5\% 
\dau (worst case) they are 8\% at 2\,\gevc and 3\% at 6\,\gevc for 
CNT-MPCS (Au-going). The uncertainty for the $-c_2/c_1$ is very similar 
to that of $c_2$ owing to a smaller uncertainty of $c_1$. This study was 
also performed for CNT-MPCN ($d$-going) correlations, obtaining 4\% 
(2\,\gevc) and 2\% (6\,\gevc) for \pp and 12\% (2\,\gevc) and 
3\,\% (6\,\gevc) for the 0\%--5\,\% \dau. Both CNT-MPCS and CNT-MPCN show 
consistent systematic uncertainties given the large statistical 
uncertainties in the CNT-MPCN correlations. Considering the better 
statistical precision for the CNT-MPCS correlations, we quoted the 
errors for them as the systematic uncertainties for the final results. 
There is a possible systematic uncertainty associated with the mixed 
event distributions $M(\Delta\phi, \pt)$. This uncertainties are 
effectively folded during the procedure of the systematic uncertainty 
estimate described above.

%%============================================
%
% Start of Results and Discussion section
%
%

\section{Results and discussions 
\label{ResultAndDiscuss}}

We present the corrected correlation functions 
[Eq.~\ref{eq:side_band_formula}], together with the four-term Fourier fit 
functions [Eq.~\ref{eq:four_fit}], across a range of collision 
systems and \piz \pt bins, for both CNT-MPCS (Au-going) and CNT-MPCN 
($d$-going) combinations. Representative samples for the bins 
$3<p_T<3.5$\,\gevc and $5<p_T<6$\,\gevc appear in 
Figs.~\ref{Fig:Pi0MPCSummary_1} and \ref{Fig:Pi0MPCSummary_2}, while the 
full sets are shown in the Appendix.

%%=================
%
% Figure with correlation functions at pT=3-3.5GeV/$c$
%

The correlation functions are largely dominated by a dipole component 
($n=1$), and higher components ($n>1$) contribute to form a near-side 
enhancement structure in the near-side ($\Delta\phi \approx 0$) of the 
functions.  The dipole component is usually attributed to the 
back-to-back dijet contribution and momentum conservation in the system. 
With the large psuedorapidity gap employed ($|\Delta\eta|>3$), the 
near-side particles of the dijet triggered by \piz ($|\eta|<0.35$) will 
not form a peak at $\Delta\phi\approx0$ in the MPCs ($3.1<|\eta|<3.9$). 
Therefore, the near-side enhancement is formed by other sources, 
possibly a quadrupole flow from a bulk medium.  The characteristic 
structure is clearly visible for CNT-MPCS (Au-going), but not for 
CNT-MPCN (d-going).  In addition, the structure is more prominent in the 
more central collisions (e.g. see the first plot in the Appendix), and 
it gradually disappears with both decreasing centrality and increasing 
\pt.  The trend in the CNT-MPCS correlation hints that the 
characteristic structure has a similar characteristic as the 
hydrodynamical particle flow in A+A collisions. Looking at the evolution 
of the individual Fourier-components $c_i$ with centrality and \pt 
provides a richer and more quantitative picture.  As seen in 
Figs.~\ref{Fig:Pi0MPCSummary_1} and \ref{Fig:Pi0MPCSummary_2}, the $c_3$ 
and $c_4$ are both very small, and are found to be consistent with zero 
within uncertainties.  Therefore, we discuss here only the centrality- and 
\pt-dependence of the dipole ($c_1$) and quadrupole ($c_2$) coefficients.

%---------------------------------------------------  Fig_7
\begin{figure*}[tbph]
\begin{minipage}[t]{0.49\linewidth}
\includegraphics[width=0.99\linewidth]{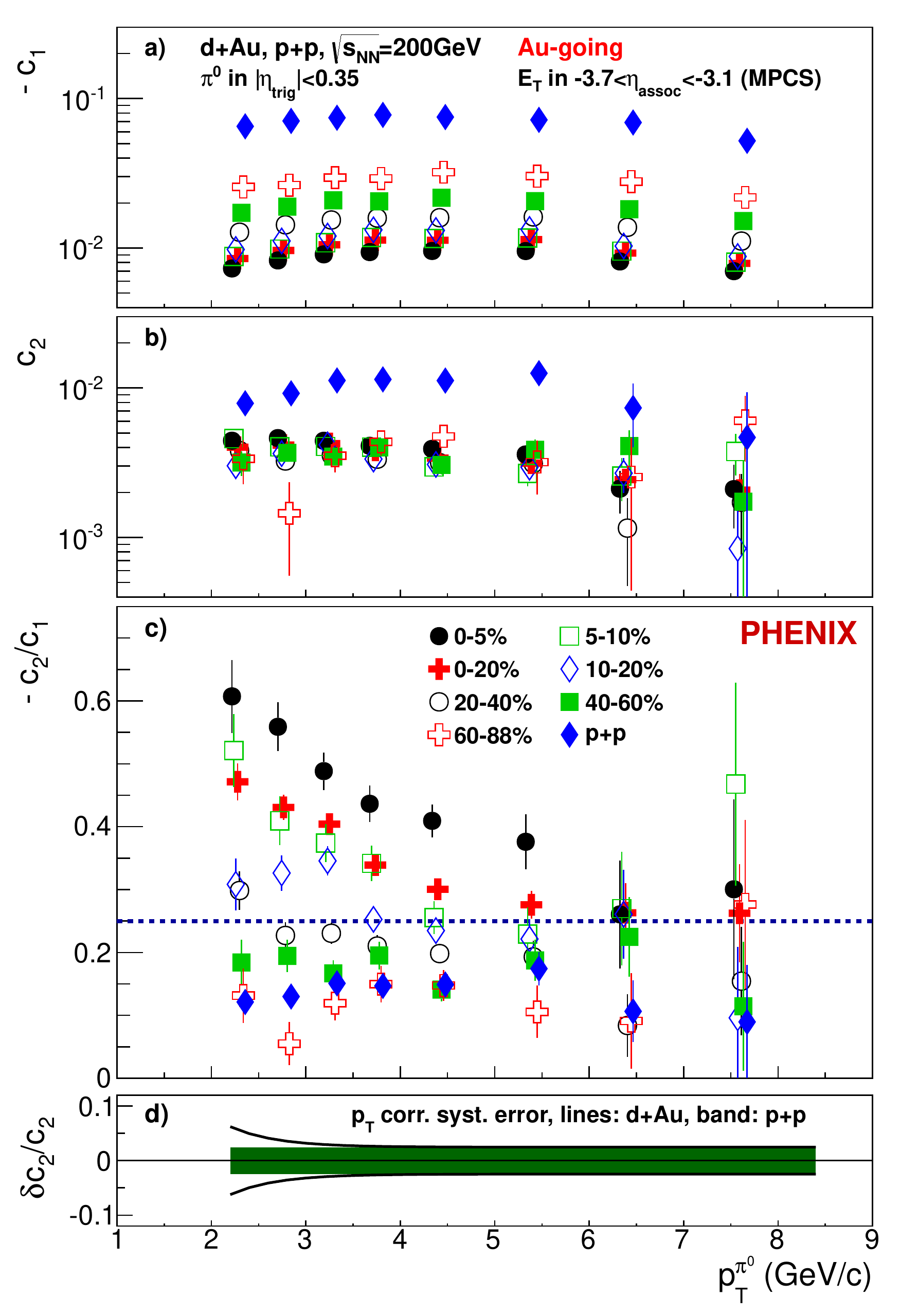}
\caption{Fourier fit coefficients for CNT-MPCS (Au-going) correlations, 
as a function of collision system and \piz \pt: (a) the negative of the 
dipole coefficient, $-c_1$; (b) the quadrupole coefficient $c_2$; (c) 
the ratio $-c_2/c_1$; (d) fractional systematic uncertainty on the 
quadrupole coefficient $c_2$ for \dau~(lines) and \pp~(band).  The 
dotted line at 0.25 in panel (c) the dashed [blue] line marks the nominal 
threshold, above which the correlation function would exhibit a 
near-side local maximum (see text).
}
\label{Fig:summary_c1c2_S}
\end{minipage}
%---------------------------------------------------  Fig_8
\begin{minipage}[t]{0.49\linewidth}
\includegraphics[width=0.99\linewidth]{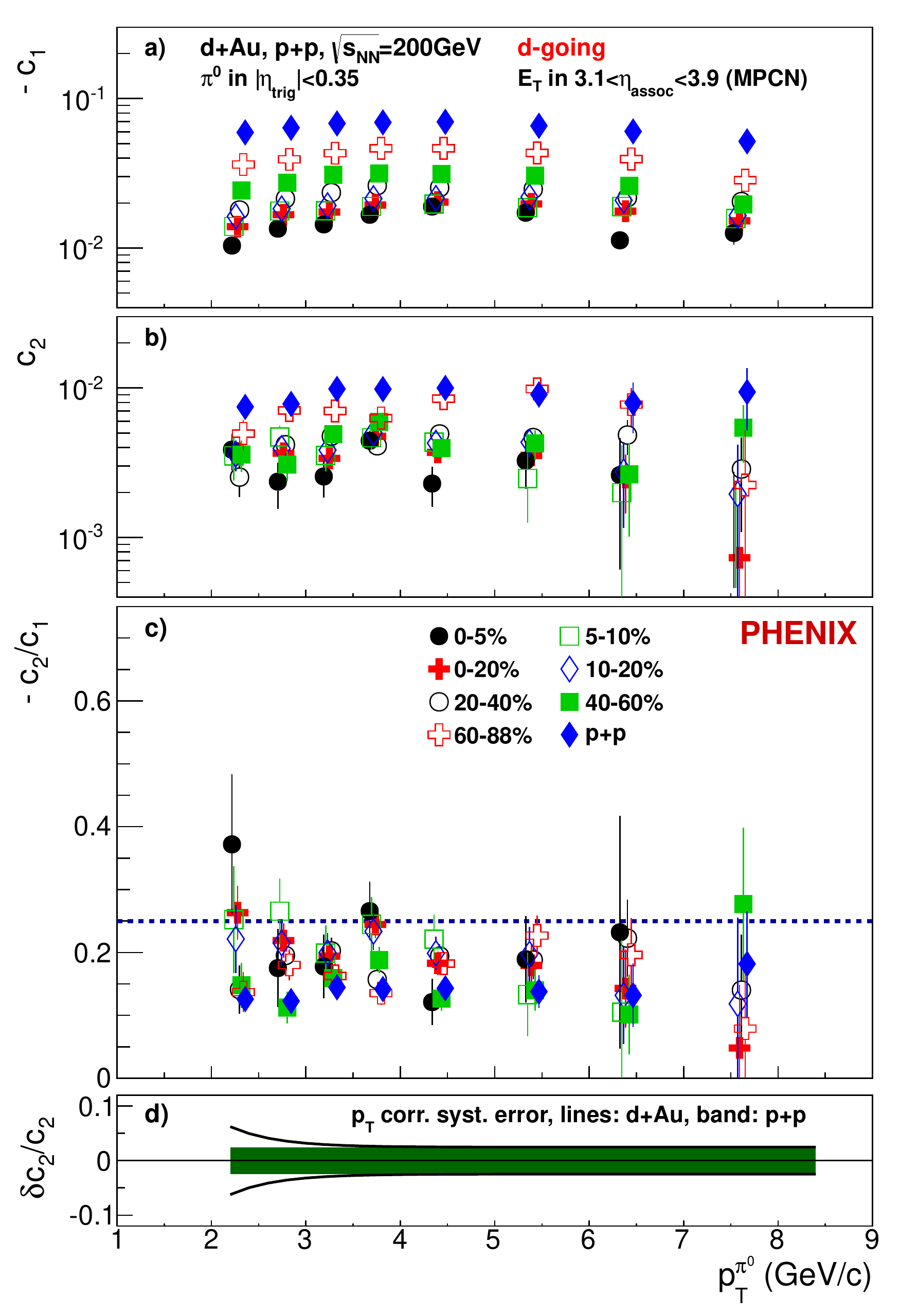}
\caption{The same as Fig.~\ref{Fig:summary_c1c2_S}, except for CNT-MPCN 
($d$-going) correlations.
}
\label{Fig:summary_c1c2_N}
\end{minipage}
\end{figure*}

% Start discussion on c_1 results
%

The $c_1$ (dipole) values for CNT-MPCS correlations are summarized in 
Fig.~\ref{Fig:summary_c1c2_S}(a).  They exhibit a definite ordering with 
system size: the largest negative values are observed in \pp, the 
smallest ones in the most central \dau. Similar ordering, albeit 
with smaller absolute differences, can be seen for CNT-MPCN in
Fig.~\ref{Fig:summary_c1c2_N}.  This trend is similar to the decrease 
of the absolute value of $c_1$ with increasing multiplicity that was 
observed in~\cite{Adamczyk:2015xjc}.  If the negative $c_1$ at large 
$\Delta\eta$ is indeed a consequence of a dijet fragmentation into the 
CNT and MPC regions, then we would expect the effect to be diluted as 
the underlying event multiplicity increases. Because the overall 
multiplicity on the $d$-going side is smaller we would also expect a 
larger magnitude for $c_1$ there compared with the Au-going direction, 
as seen in the data. Interestingly, the $c_1$ coefficients vary with \pt 
and have a maximum magnitude around 4--5\,\gevc. It may be related to 
the fact that this is the \pt region where hard scattering becomes 
dominant over bulk phenomena that govern particle production at lower 
\pt.

% Start discussion on c_2 results
%

The \pt and centrality dependence of $c_2$ (quadrupole) values in 
CNT-MPCS and CNT-MPCN correlations are shown in panel (b) of 
Figs.~\ref{Fig:summary_c1c2_S} and \ref{Fig:summary_c1c2_N}, along with 
their \pt-correlated systematic uncertainties in panel (d). For 
\pp collisions the two distributions are compatible, as expected 
for the symmetric system. The $c_2$ in \pp collisions are roughly 
double those seen in \dau (including the most peripheral bin), and 
the \pt-dependence of their magnitudes is similar to that of the $c_1$.  
For \dau the $c_2$ for CNT-MPCN and CNT-MPCS correlations are 
similar in magnitude, but with the CNT-MPCN showing a greater spread 
with centrality.  The $c_2$ are small and decreasing as a function of 
\pt, but nonvanishing in the available \pt range, proving that the 
quadrupole component is present.

To gauge the magnitude of characteristic-structure correlations as a 
measure of a bulk property of the system, we calculated $-c_2/c_1$, 
the ratio of $c_2$ (quadrupole) to $-c_1$ (dipole), for all \pp 
and \dau systems, as shown in Fig.~\ref{Fig:summary_c1c2_S}(c). For 
the CNT-MPCS correlations [Fig.~\ref{Fig:summary_c1c2_S}(a)] the data 
exhibit a well-defined ordering with system centrality, within errors, 
from the most central \dau down to the most peripheral (60\%--88\%) 
which is consistent with the \pp. We then see a smooth evolution 
from the most central collisions observed at lowest \pt, where the 
near-side correlations are most prominent and which would be 
expected to have the largest contribution from a collective source, to 
the more peripheral and higher \pt limit, where the near-side 
correlation vanishes and elementary processes are expected to dominate. 
The trend is very different for CNT-MPCN correlations 
[Fig.~\ref{Fig:summary_c1c2_N}(b)]. Here all the $-c_2/c_1$ ratios are 
consistent for both \dau and \pp collisions, indicating no 
additional near-side correlations in \dau over \pp collisions 
for any system across the entire \piz \pt range studied here. There is 
also no visible ordering of $-c_2/c_1$ with system centrality for 
$\pt>2.5$\,\gevc, in contrast to the CNT-MPCS case, within 
uncertainties.

The $c_1$ and $c_2$ for the symmetric \pp collisions are somewhat
different between CNT-MPCS and CNT-MPCN, which results from the
difference of psuedorapidity coverage in MPCN~($3.1<\eta<3.9$) versus
MPCS~($-3.7<\eta<-3.1$).  The fact that the $-c1/c2$ are very consistent
indicates that the same phenomena is observed in each direction.

%---------------------------------------------------  Fig_9
\begin{figure}[tbh]
\includegraphics[width=1.0\linewidth]{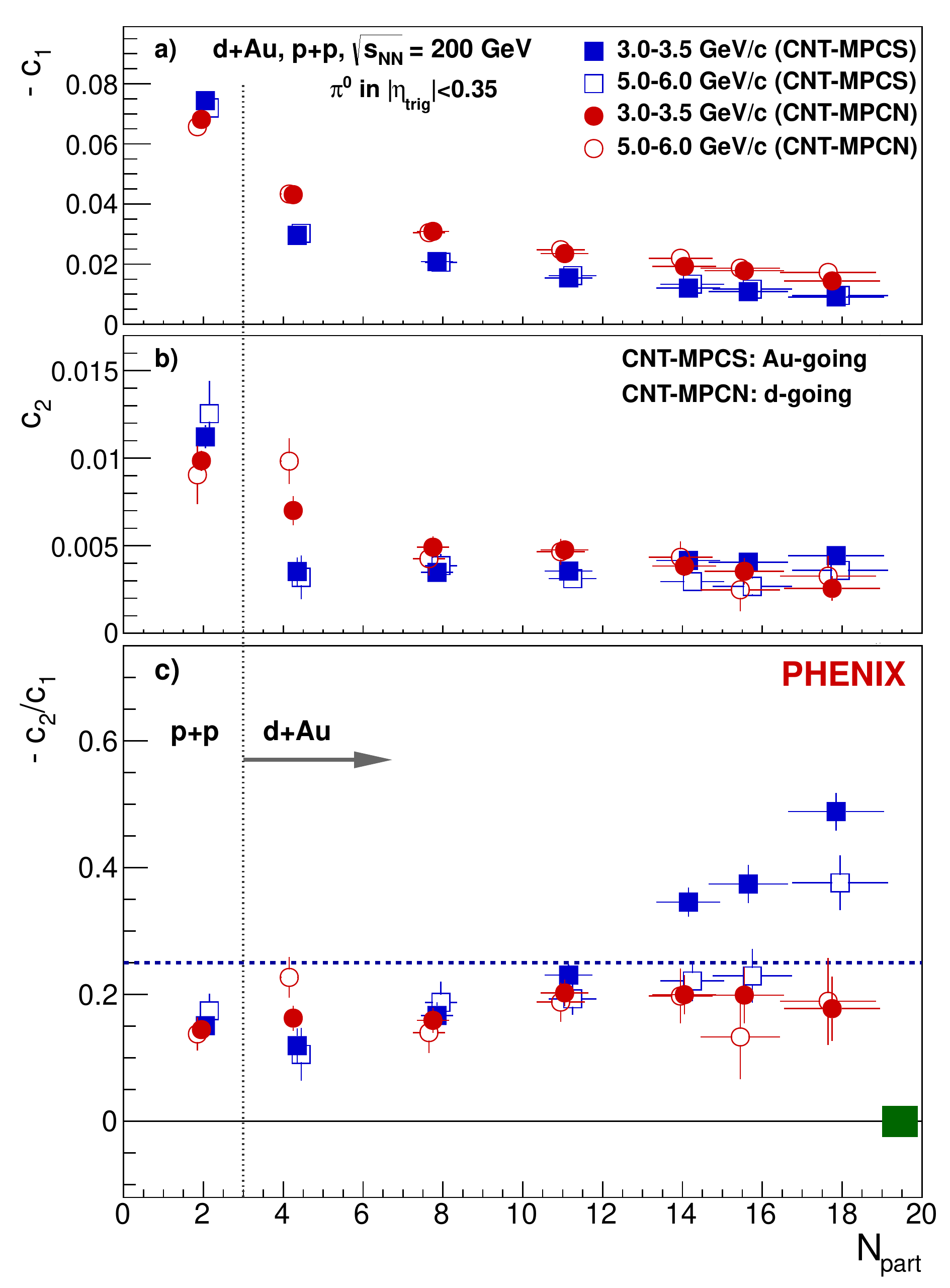}
\caption{
System centrality dependence of correlation coefficients (a) $-c_1$, (b) 
$c_2$, and (c) the ratio $-c_2/c_1$ as a function of $N_{\rm part}$. 
Results for both CNT-MPCS (Au-going direction) and CNT-MPCN ($d$-going 
direction) correlations are each shown for two \pt regions. The box 
[green] shows the systematic uncertainty on the $-c_2/c_1$ ratio, which 
is constant with $N_{\rm part}$. The uncertainties on the $N_{\rm part}$ 
estimates are shown on each data point as horizontal error bars.
}
\label{Fig:summary_NSOverCent}
\end{figure}

Recently, attempts have been made to develop methods that effectively 
subtract the nonflow contributions present in two-particle correlations, 
as measured in \pdA 
collisions~\cite{Aad:2015gqa,Aaboud:2016yar,CMS:2012qk,Khachatryan:2016ibd}. 
Despite their differences, all of these methods rely on the assumption 
that one can identify a class of events (usually \pp or peripheral \pdA) 
with low enough multiplicity such that the corresponding correlation 
function can be attributed entirely to nonflow. However, there is 
currently no consensus in the field regarding how the subtraction 
procedure should be carried out. This paper therefore focuses on the 
shape analysis of the correlation functions, leaving nonflow subtraction 
outside of the scope. However, we point out that the quantity 
$-c_2/c_1$ encodes some information about the relative strength of 
nonflow, and its comparison between collision systems can provide useful 
insight.

Another shape study of the near-side correlations can be performed by 
examining the second derivative of $dN/d(\Delta\phi)$. If we approximate 
the $n>2$ coefficients as negligible ($c_3 \approx c_4 \approx 0$) then 
the condition of having a local maximum at $\Delta\phi=0$ corresponds to

\begin{equation}
(\partial^2/\partial\Delta\phi^2)(dN/d\Delta\phi) \propto -c_1 - 4 c_2 < 0
\end{equation}

\noindent The observed positive $c_2$ and negative $c_1$ lead us to use 
the threshold of $-c_2/c_1 > 0.25$ as the condition indicating that a 
near-side correlation with a local maximum is present in the correlation 
function, as also pointed out in the literature~\cite{Trainor:2010uk}.
The dotted lines in panel (c) in 
Figs.~\ref{Fig:summary_c1c2_S} and \ref{Fig:summary_c1c2_N} indicate 
this threshold.  For the CNT-MPCS correlations the data are clearly above 
the threshold for the more central \dau collisions, out to 20\%, and 
for lower $p_T<6$\,\gevc, indicating that the shapes have a local 
maximum.  For the CNT-MPCN correlations, all the $-c_2/c_1$ ratios 
consistently lie below 0.25 for both \dau and \pp collisions, 
indicating no local maximum. It should be noted that the absence of a 
local maximum doesn't necessarily imply that the near-side contribution 
is absent.

We now examine the system and centrality dependence of the correlation 
functions.  Figure~\ref{Fig:summary_NSOverCent} shows the $c_1$, $c_2$, 
and $-c_2/c_1$ as a function of the mean number of collision 
participants $N_{\rm part}$~\cite{Adare:2013nff} for the two selected 
\pt ranges 3--3.5\,\gevc and 5--6\,\gevc.

The values for both CNT-MPCS and CNT-MPCN are shown. The smooth decrease of 
$c_1$ with $N_{\rm part}$ is clearly seen for both \pt selections, but 
the decrease of $c_1$ for the CNT-MPCS is more rapid compared to that of 
CNT-MPCN.  In contrast, the $c_2$ is flat or exhibits little 
increase (decrease) as a function of $N_{\rm part}$ for CNT-MPCN (CNT-MPCS) 
correlations, except for the lowest $N_{\rm part}$.  In $-c_2/c_1$, 
where individual $-c_1$ and $c_2$ trends are combined, the data for 
CNT-MPCS show a smooth rising trend, stronger for the lower \pt 
selection, while the $-c_2/c_1$ for CNT-MPCN correlations displays no 
evolution with $N_{\rm part}$ at all from \pp to the most central 
\dau collisions.  This observation clearly shows again that the 
characteristic structure is clearly seen in Au-going direction, rather than 
in d-going direction, and ceases at high \pt which is a similar 
characteristic as the hydrodynamical particle flow in $A$$+$$A$ collisions.

The centrality dependence of $-c_1/c_2$ can be understood in terms of 
the asymmetry of the charged particle pseudorapidity distributions with 
respect to $\eta=0$ in \dau collisions~\cite{Back:2004mr}. When going to 
greater centrality, the results indicate that the characteristic 
structure is shifted in the Au-going direction, similar to the 
charged-particle pseudorapidity distributions. This is consistent with 
the findings of the STAR experiment~\cite{Adamczyk:2015xjc} in the 
region where the \pt ranges overlap.  There is a possible fluctuation of 
the event plane as a function of psuedorapidity as observed by the CMS 
experiment at the LHC~\cite{Khachatryan:2015oea}.  Although this may 
partly explain the centrality-dependent difference between CNT-MPCN and 
CNT-MPCS, our measurements lack the precision to gauge the effect.  
These results provide a strong argument for studying long-range 
correlations in asymmetric systems separately in the forward/backward 
directions.

\section{Summary}

We have measured long-range azimuthal correlations between 
high-transverse-momentum ($2<p_T<11$\,\gevc) \piz observed at 
midrapidity ($|\eta|<0.35$) and particles produced either at forward 
($3.1<\eta<3.9$) and backward ($-3.7<\eta<-3.1$) rapidity in \dau 
and \pp collisions at \snn=200\,GeV. The centrality and 
\pt-dependent two-particle correlations were fitted with a 
Fourier-series up to the fourth term.  While the 3rd and 4th 
coefficients ($c_3, c_4$) were consistent with zero within uncertainties, 
the $c_1$ (dipole) values exhibit a definite ordering with the system 
size both in the Au-going and $d$-going directions. The $c_2$ 
(quadrupole) values exhibit similar magnitudes for both directions. 
However, $-c_2/c_1$ values exhibit well-defined ordering with system 
centrality and decrease with increasing \pt in the Au-going direction, while 
the values are consistent over all systems and \pt in the $d$-going 
direction. This implies that the characteristic structure clearly exists in 
the Au-going direction, rather than in the $d$-going direction, and 
ceases at high \pt which is a similar characteristic as the 
hydrodynamical particle flow in $A$$+$$A$ collisions. The difference of the 
behavior in the Au-going and the $d$-going direction can be understood 
from the fact that the characteristic structure is shifted in the Au-going 
direction toward more central collisions, similar to the 
charged-particle pseudorapidity distributions.  This suggests that 
looking at two directions in asymmetric systems is essential.

%%%%%%%%%%%%%%%%%%%%%%%%%  Acknowledgements 

\section*{ACKNOWLEDGMENTS}   % Run-17 long form for all journals

We thank the staff of the Collider-Accelerator and Physics
Departments at Brookhaven National Laboratory and the staff of
the other PHENIX participating institutions for their vital
contributions.  We acknowledge support from the 
Office of Nuclear Physics in the
Office of Science of the Department of Energy,
the National Science Foundation, 
Abilene Christian University Research Council, 
Research Foundation of SUNY, and
Dean of the College of Arts and Sciences, Vanderbilt University 
(U.S.A),
Ministry of Education, Culture, Sports, Science, and Technology
and the Japan Society for the Promotion of Science (Japan),
Conselho Nacional de Desenvolvimento Cient\'{\i}fico e
Tecnol{\'o}gico and Funda\c c{\~a}o de Amparo {\`a} Pesquisa do
Estado de S{\~a}o Paulo (Brazil),
Natural Science Foundation of China (People's Republic of China),
Croatian Science Foundation and
Ministry of Science and Education (Croatia),
Ministry of Education, Youth and Sports (Czech Republic),
Centre National de la Recherche Scientifique, Commissariat
{\`a} l'{\'E}nergie Atomique, and Institut National de Physique
Nucl{\'e}aire et de Physique des Particules (France),
Bundesministerium f\"ur Bildung und Forschung, Deutscher
Akademischer Austausch Dienst, and Alexander von Humboldt Stiftung (Germany),
J. Bolyai Research Scholarship, EFOP, the New National Excellence
Program ({\'U}NKP), NKFIH, and OTKA (Hungary),
Department of Atomic Energy and Department of Science and Technology (India), 
Israel Science Foundation (Israel), 
Basic Science Research Program through NRF of the Ministry of Education (Korea),
Physics Department, Lahore University of Management Sciences (Pakistan),
Ministry of Education and Science, Russian Academy of Sciences,
Federal Agency of Atomic Energy (Russia),
VR and Wallenberg Foundation (Sweden), 
the U.S. Civilian Research and Development Foundation for the
Independent States of the Former Soviet Union, 
the Hungarian American Enterprise Scholarship Fund,
the US-Hungarian Fulbright Foundation,
and the US-Israel Binational Science Foundation.

%%%%%%%%%%%%%%%%%%%%%%%%%  Appendix

\section*{APPENDIX}

Figures~\ref{Fig:Pi0MPCSouth1}--\ref{Fig:Pi0MPCNorth2} show data points 
of the normalized correlation functions in CNT-MPCS and CNT-MPCN for all 
\dau centralities and in \pt bins of the trigger \piz in CNT 
($|\eta_{\rm trig}|<$0.35), along with the fitted Fourier-components and 
their sum.  Note the changes in $y$-scale from 
Figs.~\ref{Fig:Pi0MPCSouth1} and \ref{Fig:Pi0MPCNorth1} to 
Figs.~\ref{Fig:Pi0MPCSouth2} and \ref{Fig:Pi0MPCNorth2}.
Although the correlation functions are shown up to \pt=11\,\gevc, it is 
clear that the statistical precision is poor for the 9--11\,\gevc data.  
Therefore, the $c_1$, $c_2$ and $-c_2/c_1$ in this paper are shown 
only up to 9\,\gevc.

\begin{turnpage}
%---------------------------------------------------  Fig_10
\begin{figure*}[th]
\includegraphics[width=0.99\linewidth]{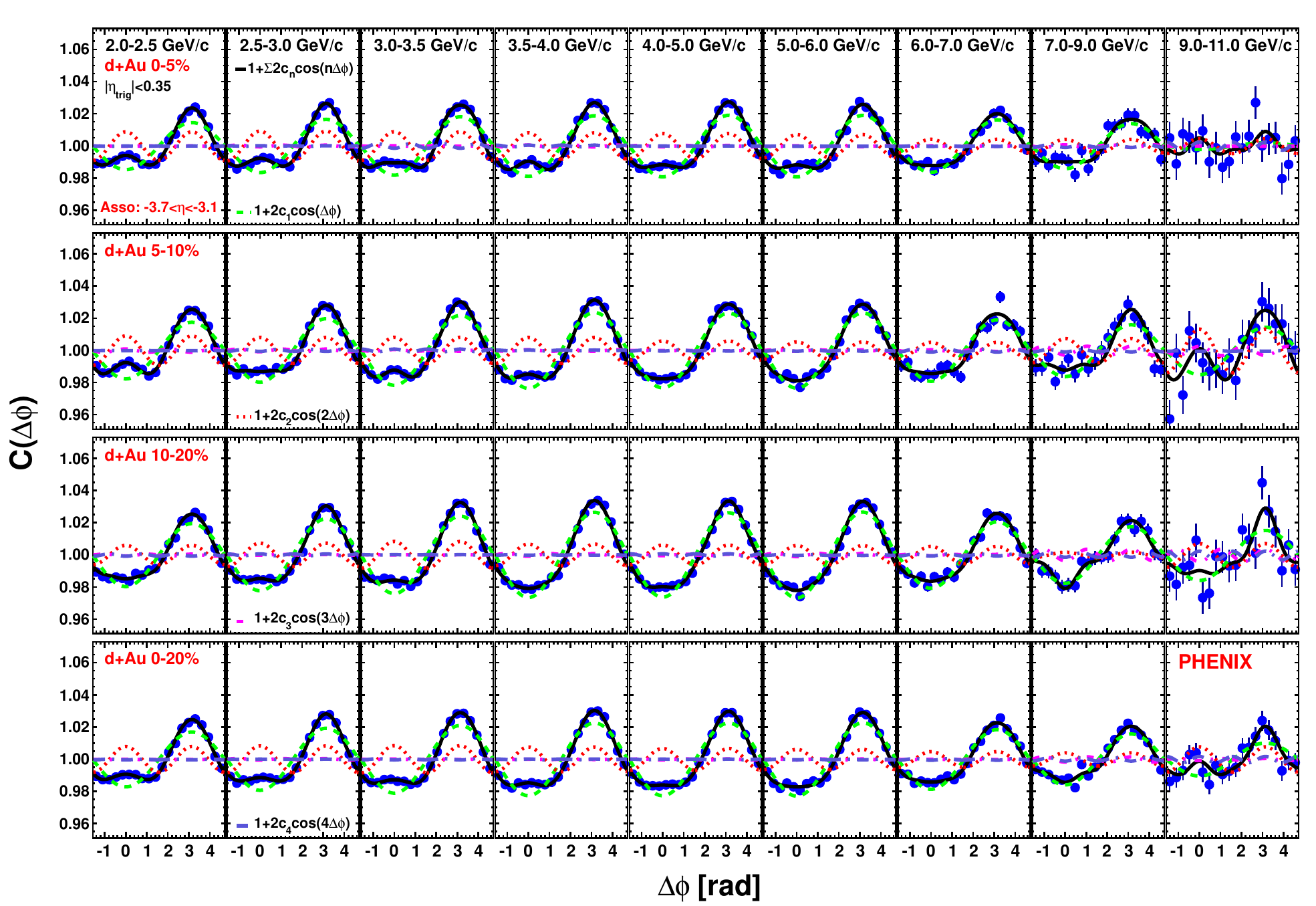}
\caption{CNT-MPCS correlation functions for 0\%--5\%, 5\%--10\%, 
10\%--20\%, 0\%--20\% \dau collisions for 2.0$<p_T<$11\,\gevc.
}
\label{Fig:Pi0MPCSouth1}
\end{figure*}

%---------------------------------------------------  Fig_11
\begin{figure*}[th]
\includegraphics[width=0.99\linewidth]{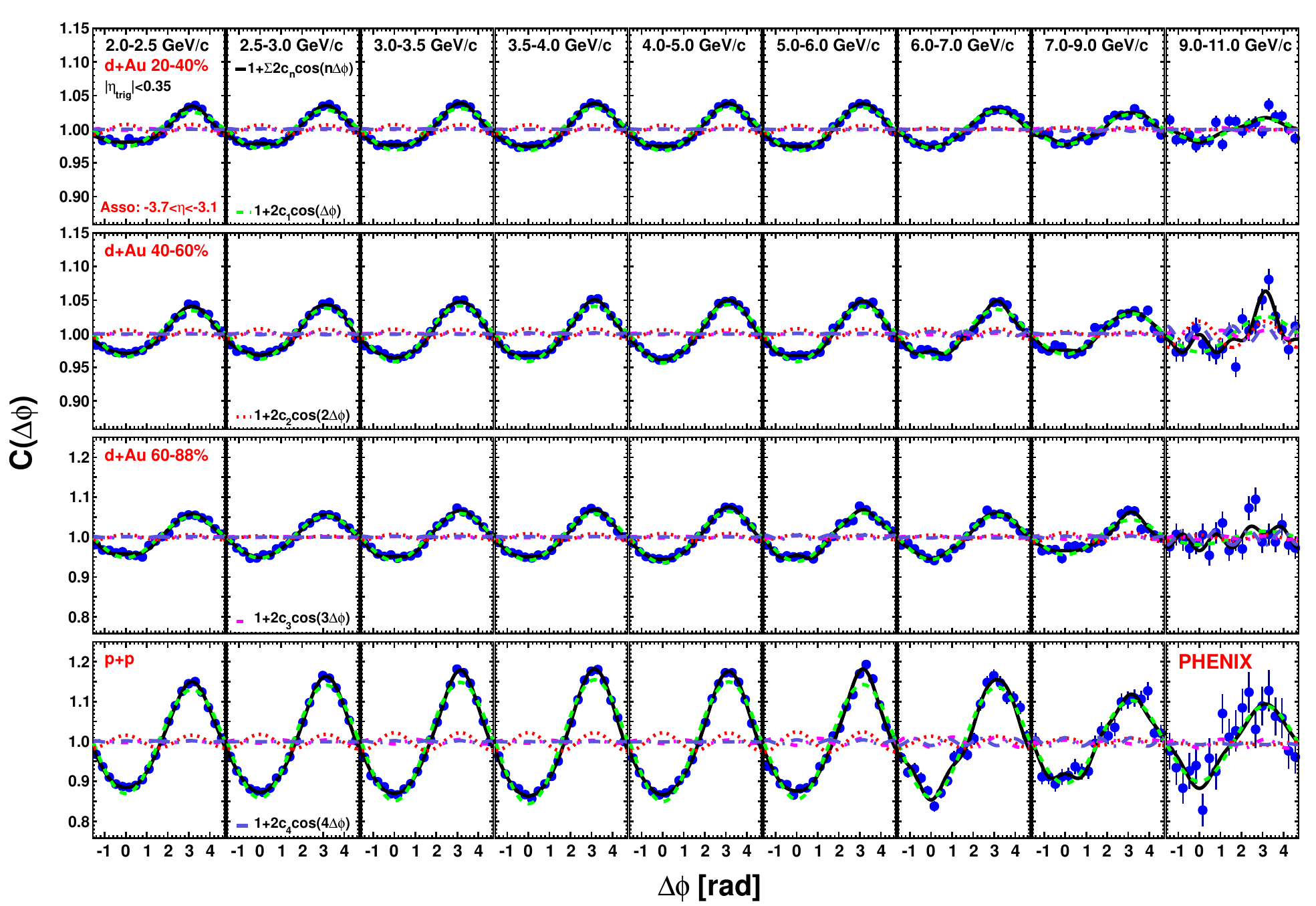}
\caption{CNT-MPCS correlation functions for 20\%--40\%, 40\%--60\%, 
60\%--88\% \dau collisions and \pp collisions for 
$2.0<p_T<11$\,\gevc.
}
\label{Fig:Pi0MPCSouth2}
\end{figure*}

%---------------------------------------------------  Fig_12
\begin{figure*}[th]
\includegraphics[width=0.99\linewidth]{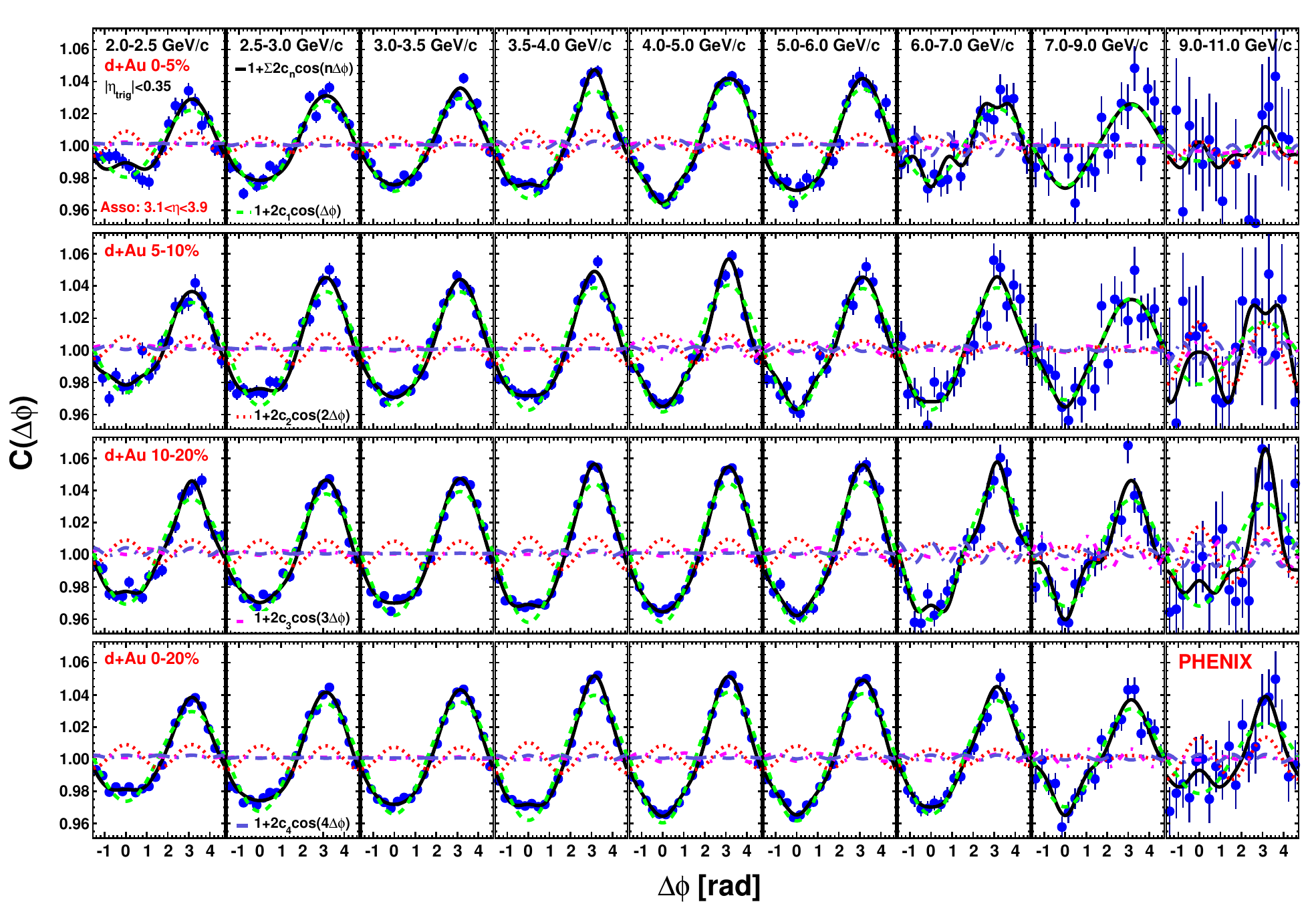}
\caption{CNT-MPCN correlation functions for 0\%--5\%, 5\%--10\%, 
10\%--20\%, 0\%--20\% \dau collisions for $2.0<p_T<11$\,\gevc.
}
\label{Fig:Pi0MPCNorth1}
\end{figure*}

%---------------------------------------------------  Fig_13
\begin{figure*}[th]
\includegraphics[width=0.99\linewidth]{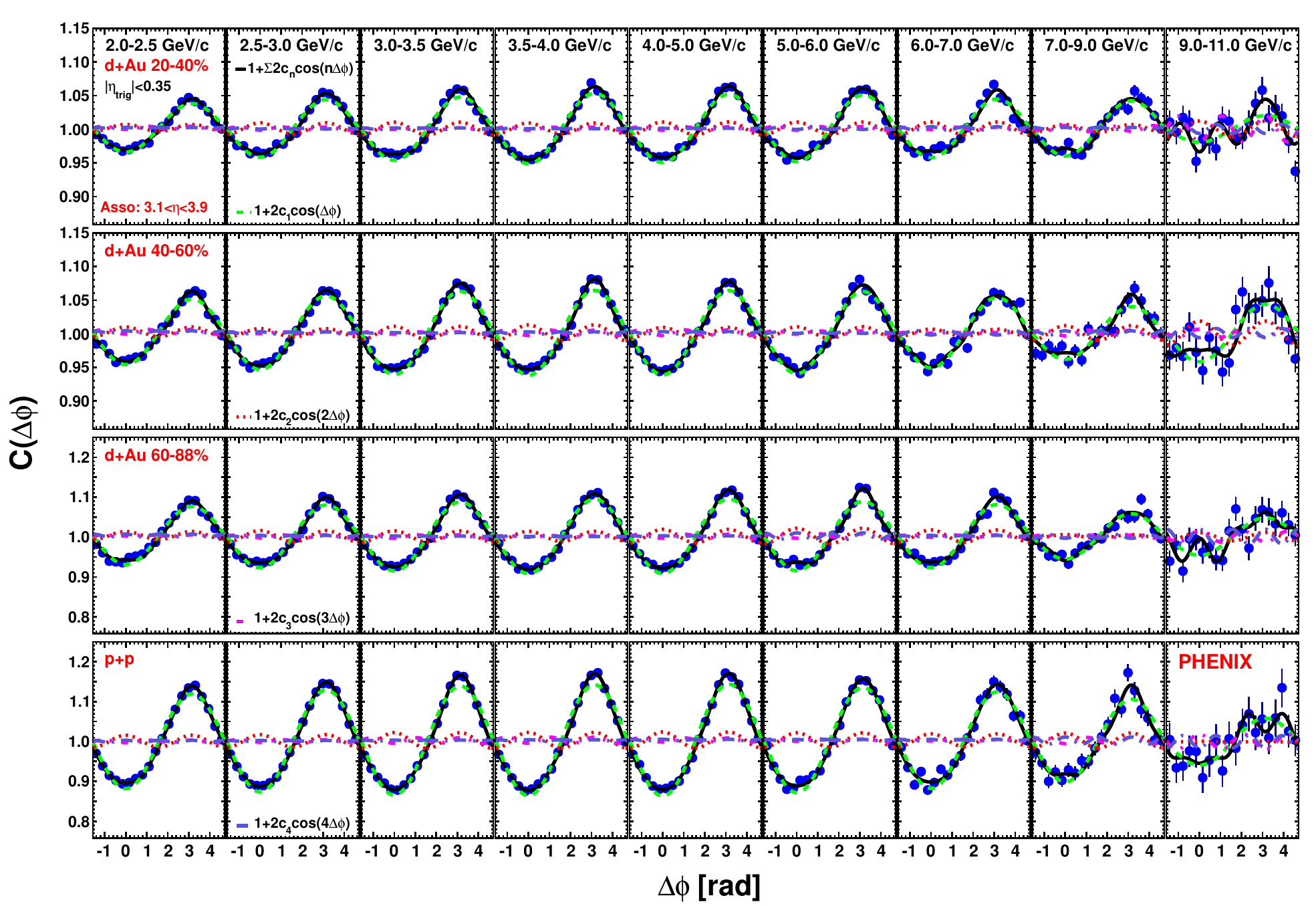}
\caption{CNT-MPCN correlation functions for 20\%--40\%, 40\%--60\%, 
60\%--88\% \dau collisions and \pp collisions for 
$2.0<p_T<11$\,\gevc.
}
\label{Fig:Pi0MPCNorth2}
\end{figure*}
\end{turnpage}

%\appendix
%\setcounter{table}{0} \renewcommand{\thetable}{A.\arabic{table}} 

\clearpage

%%%%%%%%%%%%%%%%%%%%%%%%%%%  References 

%\bibliography{ppg192x2} 

%merlin.mbs apsrev4-1.bst 2010-07-25 4.21a (PWD, AO, DPC) hacked
%Control: key (0)
%Control: author (0) dotless jnrlst
%Control: editor formatted (1) identically to author
%Control: production of article title (0) allowed
%Control: page (1) range
%Control: year (0) verbatim
%Control: production of eprint (0) enabled
%
 
\end{document}